\documentstyle[11pt]{article}

\newcommand{\stln}{\setlength{\unitlength}{2.2ex}}
\newcommand{\fr}{\framebox(1,1){}}
\newcommand{\sfr}{\framebox(1,1){\begin{picture}(1,1)
  \put(0,0){\line(1,1){1}}\end{picture}}}

\newcommand{\onebox}
{\stln \lower1.4ex\hbox{
\begin{picture}(1.6,1.6)
\put(.3,.3){\fr}
\end{picture}}}

\newcommand{\twobox}
{\stln \lower1.4ex\hbox{
\begin{picture}(2.6,1.6)
\put(.3,.3){\fr}
\put(1.3,.3){\fr}
\end{picture}}}

\newcommand{\threebox}
{\stln \lower1.4ex\hbox{
\begin{picture}(3.6,1.6)
\multiput(.3,.3)(1,0){3}{\fr}
\end{picture}}}

\newcommand{\fourbox}
{\stln \lower1.4ex\hbox{
\begin{picture}(4.6,1.6)
\multiput(.3,.3)(1,0){4}{\fr}
\end{picture}}}

\newcommand{\fivebox}
{\stln \lower1.4ex\hbox{
\begin{picture}(5.6,1.6)
\multiput(.3,.3)(1,0){5}{\fr}
\end{picture}}}

\newcommand{\sixbox}
{\stln \lower1.4ex\hbox{
\begin{picture}(6.6,1.6)
\multiput(.3,.3)(1,0){6}{\fr}
\end{picture}}}

\newcommand{\genrowbox}
{\stln \lower1.4ex\hbox{
\begin{picture}(7.6,1.6)
\multiput(.3,.3)(1,0){3}{\fr}
\put(3.3,.3){\framebox(3,1){$\cdots$}}
\put(6.3,.3){\fr}
\end{picture}}}

\newcommand{\oneonebox}
{\stln \lower2.6ex\hbox{
\begin{picture}(1.6,2.6)
\put(.3,.3){\fr}
\put(.3,1.3){\fr}
\end{picture}}}

\newcommand{\twoonebox}
{\stln \lower2.6ex\hbox{
\begin{picture}(2.6,2.6)
\put(.3,1.3){\fr}
\put(1.3,1.3){\fr}
\put(0.3,0.3){\fr}
\end{picture}}}

\newcommand{\threeonebox}
{\stln \lower2.6ex\hbox{
\begin{picture}(3.6,2.6)
\multiput(.3,1.3)(1,0){3}{\fr}
\put(.3,.3){\fr}
\end{picture}}}

\newcommand{\fouronebox}
{\stln \lower2.6ex\hbox{
\begin{picture}(4.6,2.6)
\multiput(.3,1.3)(1,0){4}{\fr}
\put(.3,.3){\fr}
\end{picture}}}

\newcommand{\fiveonebox}
{\stln \lower2.6ex\hbox{
\begin{picture}(5.6,2.6)
\multiput(.3,1.3)(1,0){5}{\fr}
\put(.3,.3){\fr}
\end{picture}}}

\newcommand{\sixonebox}
{\stln \lower2.6ex\hbox{
\begin{picture}(6.6,2.6)
\multiput(.3,1.3)(1,0){6}{\fr}
\put(.3,.3){\fr}
\end{picture}}}

\newcommand{\twotwobox}
{\stln \lower2.6ex\hbox{
\begin{picture}(2.6,2.6)
\put(.3,.3){\fr}
\put(.3,1.3){\fr}
\put(1.3,.3){\fr}
\put(1.3,1.3){\fr}
\end{picture}}}

\newcommand{\threetwobox}
{\stln \lower2.6ex\hbox{
\begin{picture}(3.6,2.6)
\multiput(.3,1.3)(1,0){3}{\fr}
\put(.3,.3){\fr}
\put(1.3,.3){\fr}
\end{picture}}}

\newcommand{\fourtwobox}
{\stln \lower2.6ex\hbox{
\begin{picture}(4.6,2.6)
\multiput(.3,1.3)(1,0){4}{\fr}
\put(.3,.3){\fr}
\put(1.3,.3){\fr}
\end{picture}}}

\newcommand{\fivetwobox}
{\stln \lower2.6ex\hbox{
\begin{picture}(5.6,2.6)
\multiput(.3,1.3)(1,0){5}{\fr}
\put(.3,.3){\fr}
\put(1.3,.3){\fr}
\end{picture}}}

\newcommand{\sixtwobox}
{\stln \lower2.6ex\hbox{
\begin{picture}(6.6,2.6)
\multiput(.3,1.3)(1,0){6}{\fr}
\put(.3,.3){\fr}
\put(1.3,.3){\fr}
\end{picture}}}

\newcommand{\threethreebox}
{\stln \lower2.6ex\hbox{
\begin{picture}(3.6,2.6)
\multiput(.3,1.3)(1,0){3}{\fr}
\multiput(.3,.3)(1,0){3}{\fr}
\end{picture}}}

\newcommand{\fourthreebox}
{\stln \lower2.6ex\hbox{
\begin{picture}(4.6,2.6)
\multiput(.3,1.3)(1,0){4}{\fr}
\multiput(.3,.3)(1,0){3}{\fr}
\end{picture}}}

\newcommand{\fivethreebox}
{\stln \lower2.6ex\hbox{
\begin{picture}(5.6,2.6)
\multiput(.3,1.3)(1,0){5}{\fr}
\multiput(.3,.3)(1,0){3}{\fr}
\end{picture}}}

\newcommand{\sixthreebox}
{\stln \lower2.6ex\hbox{
\begin{picture}(6.6,2.6)
\multiput(.3,1.3)(1,0){6}{\fr}
\multiput(.3,.3)(1,0){3}{\fr}
\end{picture}}}

\newcommand{\fourfourbox}
{\stln \lower2.6ex\hbox{
\begin{picture}(4.6,2.6)
\multiput(.3,1.3)(1,0){4}{\fr}
\multiput(.3,.3)(1,0){4}{\fr}
\end{picture}}}

\newcommand{\fivefourbox}
{\stln \lower2.6ex\hbox{
\begin{picture}(5.6,2.6)
\multiput(.3,1.3)(1,0){5}{\fr}
\multiput(.3,.3)(1,0){4}{\fr}
\end{picture}}}

\newcommand{\sixfourbox}
{\stln \lower2.6ex\hbox{
\begin{picture}(6.6,2.6)
\multiput(.3,1.3)(1,0){6}{\fr}
\multiput(.3,.3)(1,0){4}{\fr}
\end{picture}}}

\newcommand{\fivefivebox}
{\stln \lower2.6ex\hbox{
\begin{picture}(5.6,2.6)
\multiput(.3,1.3)(1,0){5}{\fr}
\multiput(.3,.3)(1,0){5}{\fr}
\end{picture}}}

\newcommand{\sixfivebox}
{\stln \lower2.6ex\hbox{
\begin{picture}(6.6,2.6)
\multiput(.3,1.3)(1,0){6}{\fr}
\multiput(.3,.3)(1,0){5}{\fr}
\end{picture}}}

\newcommand{\sixsixbox}
{\stln \lower2.6ex\hbox{
\begin{picture}(6.6,2.6)
\multiput(.3,1.3)(1,0){6}{\fr}
\multiput(.3,.3)(1,0){6}{\fr}
\end{picture}}}

\newcommand{\oneoneonebox}
{\stln \lower3.8ex\hbox{
\begin{picture}(1.6,3.6)
\multiput(.3,.3)(0,1){3}{\fr}
\end{picture}}}

\newcommand{\twooneonebox}
{\stln \lower3.8ex\hbox{
\begin{picture}(2.6,3.6)
\multiput(.3,.3)(0,1){3}{\fr}
\put(1.3,2.3){\fr}
\end{picture}}}

\newcommand{\twotwoonebox}
{\stln \lower3.8ex\hbox{
\begin{picture}(2.6,3.6)
\multiput(.3,.3)(0,1){3}{\fr}
\put(1.3,1.3){\fr}
\put(1.3,2.3){\fr}
\end{picture}}}

\newcommand{\twotwotwobox}
{\stln \lower3.8ex\hbox{
\begin{picture}(2.6,3.6)
\multiput(.3,.3)(0,1){3}{\fr}
\multiput(1.3,.3)(0,1){3}{\fr}
\end{picture}}}

\newcommand{\threeoneonebox}
{\stln \lower3.8ex\hbox{
\begin{picture}(3.6,3.6)
\multiput(.3,.3)(0,1){3}{\fr}
\put(1.3,2.3){\fr}
\put(2.3,2.3){\fr}
\end{picture}}}

\newcommand{\threetwoonebox}
{\stln \lower3.8ex\hbox{
\begin{picture}(3.6,3.6)
\multiput(.3,.3)(0,1){3}{\fr}
\put(1.3,2.3){\fr}
\put(2.3,2.3){\fr}
\put(1.3,1.3){\fr}
\end{picture}}}

\newcommand{\threetwotwobox}
{\stln \lower3.8ex\hbox{
\begin{picture}(3.6,3.6)
\multiput(.3,.3)(0,1){3}{\fr}
\multiput(1.3,.3)(0,1){3}{\fr}
\put(2.3,2.3){\fr}
\end{picture}}}

\newcommand{\threethreeonebox}
{\stln \lower3.8ex\hbox{
\begin{picture}(3.6,3.6)
\multiput(.3,2.3)(1,0){3}{\fr}
\multiput(.3,1.3)(1,0){3}{\fr}
\put(.3,.3){\fr}
\end{picture}}}

\newcommand{\threethreetwobox}
{\stln \lower3.8ex\hbox{
\begin{picture}(3.6,3.6)
\multiput(.3,2.3)(1,0){3}{\fr}
\multiput(.3,1.3)(1,0){3}{\fr}
\put(.3,.3){\fr}
\put(1.3,.3){\fr}
\end{picture}}}

\newcommand{\threethreethreebox}
{\stln \lower3.8ex\hbox{
\begin{picture}(3.6,3.6)
\multiput(.3,2.3)(1,0){3}{\fr}
\multiput(.3,1.3)(1,0){3}{\fr}
\multiput(.3,.3)(1,0){3}{\fr}
\end{picture}}}

\newcommand{\gencolbox}
{\stln \lower8.6ex\hbox{
\begin{picture}(1.6,7.6)
\multiput(.3,4.3)(0,1){3}{\fr}
\put(.3,1.3){\framebox(1,3){$\vdots$}}
\put(.3,.3){\fr}
\end{picture}}}

\newcommand{\sonebox}
{\stln \lower1.4ex\hbox{
\begin{picture}(1.6,1.6)
\put(.3,.3){\sfr}
\end{picture}}}

\newcommand{\stwobox}
{\stln \lower1.4ex\hbox{
\begin{picture}(2.6,1.6)
\put(.3,.3){\sfr}
\put(1.3,.3){\sfr}
\end{picture}}}

\newcommand{\sthreebox}
{\stln \lower1.4ex\hbox{
\begin{picture}(3.6,1.6)
\multiput(.3,.3)(1,0){3}{\sfr}
\end{picture}}}

\newcommand{\sfourbox}
{\stln \lower1.4ex\hbox{
\begin{picture}(4.6,1.6)
\multiput(.3,.3)(1,0){4}{\sfr}
\end{picture}}}

\newcommand{\sfivebox}
{\stln \lower1.4ex\hbox{
\begin{picture}(5.6,1.6)
\multiput(.3,.3)(1,0){5}{\sfr}
\end{picture}}}

\newcommand{\ssixbox}
{\stln \lower1.4ex\hbox{
\begin{picture}(6.6,1.6)
\multiput(.3,.3)(1,0){6}{\sfr}
\end{picture}}}

\newcommand{\sgenrowbox}
{\stln \lower1.4ex\hbox{
\begin{picture}(7.6,1.6)
\multiput(.3,.3)(1,0){3}{\sfr}
\put(3.3,.3){\framebox(3,1){$\cdots$}}
\put(6.3,.3){\sfr}
\end{picture}}}

\newcommand{\soneonebox}
{\stln \lower2.6ex\hbox{
\begin{picture}(1.6,2.6)
\put(.3,.3){\sfr}
\put(.3,1.3){\sfr}
\end{picture}}}

\newcommand{\stwoonebox}
{\stln \lower2.6ex\hbox{
\begin{picture}(2.6,2.6)
\put(.3,1.3){\sfr}
\put(1.3,1.3){\sfr}
\put(0.3,0.3){\sfr}
\end{picture}}}

\newcommand{\sthreeonebox}
{\stln \lower2.6ex\hbox{
\begin{picture}(3.6,2.6)
\multiput(.3,1.3)(1,0){3}{\sfr}
\put(.3,.3){\sfr}
\end{picture}}}

\newcommand{\sfouronebox}
{\stln \lower2.6ex\hbox{
\begin{picture}(4.6,2.6)
\multiput(.3,1.3)(1,0){4}{\sfr}
\put(.3,.3){\sfr}
\end{picture}}}

\newcommand{\sfiveonebox}
{\stln \lower2.6ex\hbox{
\begin{picture}(5.6,2.6)
\multiput(.3,1.3)(1,0){5}{\sfr}
\put(.3,.3){\sfr}
\end{picture}}}

\newcommand{\ssixonebox}
{\stln \lower2.6ex\hbox{
\begin{picture}(6.6,2.6)
\multiput(.3,1.3)(1,0){6}{\sfr}
\put(.3,.3){\sfr}
\end{picture}}}

\newcommand{\stwotwobox}
{\stln \lower2.6ex\hbox{
\begin{picture}(2.6,2.6)
\put(.3,.3){\sfr}
\put(.3,1.3){\sfr}
\put(1.3,.3){\sfr}
\put(1.3,1.3){\sfr}
\end{picture}}}

\newcommand{\sthreetwobox}
{\stln \lower2.6ex\hbox{
\begin{picture}(3.6,2.6)
\multiput(.3,1.3)(1,0){3}{\sfr}
\put(.3,.3){\sfr}
\put(1.3,.3){\sfr}
\end{picture}}}

\newcommand{\sfourtwobox}
{\stln \lower2.6ex\hbox{
\begin{picture}(4.6,2.6)
\multiput(.3,1.3)(1,0){4}{\sfr}
\put(.3,.3){\sfr}
\put(1.3,.3){\sfr}
\end{picture}}}

\newcommand{\sfivetwobox}
{\stln \lower2.6ex\hbox{
\begin{picture}(5.6,2.6)
\multiput(.3,1.3)(1,0){5}{\sfr}
\put(.3,.3){\sfr}
\put(1.3,.3){\sfr}
\end{picture}}}

\newcommand{\ssixtwobox}
{\stln \lower2.6ex\hbox{
\begin{picture}(6.6,2.6)
\multiput(.3,1.3)(1,0){6}{\sfr}
\put(.3,.3){\sfr}
\put(1.3,.3){\sfr}
\end{picture}}}

\newcommand{\sthreethreebox}
{\stln \lower2.6ex\hbox{
\begin{picture}(3.6,2.6)
\multiput(.3,1.3)(1,0){3}{\sfr}
\multiput(.3,.3)(1,0){3}{\sfr}
\end{picture}}}

\newcommand{\sfourthreebox}
{\stln \lower2.6ex\hbox{
\begin{picture}(4.6,2.6)
\multiput(.3,1.3)(1,0){4}{\sfr}
\multiput(.3,.3)(1,0){3}{\sfr}
\end{picture}}}

\newcommand{\sfivethreebox}
{\stln \lower2.6ex\hbox{
\begin{picture}(5.6,2.6)
\multiput(.3,1.3)(1,0){5}{\sfr}
\multiput(.3,.3)(1,0){3}{\sfr}
\end{picture}}}

\newcommand{\ssixthreebox}
{\stln \lower2.6ex\hbox{
\begin{picture}(6.6,2.6)
\multiput(.3,1.3)(1,0){6}{\sfr}
\multiput(.3,.3)(1,0){3}{\sfr}
\end{picture}}}

\newcommand{\sfourfourbox}
{\stln \lower2.6ex\hbox{
\begin{picture}(4.6,2.6)
\multiput(.3,1.3)(1,0){4}{\sfr}
\multiput(.3,.3)(1,0){4}{\sfr}
\end{picture}}}

\newcommand{\sfivefourbox}
{\stln \lower2.6ex\hbox{
\begin{picture}(5.6,2.6)
\multiput(.3,1.3)(1,0){5}{\sfr}
\multiput(.3,.3)(1,0){4}{\sfr}
\end{picture}}}

\newcommand{\ssixfourbox}
{\stln \lower2.6ex\hbox{
\begin{picture}(6.6,2.6)
\multiput(.3,1.3)(1,0){6}{\sfr}
\multiput(.3,.3)(1,0){4}{\sfr}
\end{picture}}}

\newcommand{\sfivefivebox}
{\stln \lower2.6ex\hbox{
\begin{picture}(5.6,2.6)
\multiput(.3,1.3)(1,0){5}{\sfr}
\multiput(.3,.3)(1,0){5}{\sfr}
\end{picture}}}

\newcommand{\ssixfivebox}
{\stln \lower2.6ex\hbox{
\begin{picture}(6.6,2.6)
\multiput(.3,1.3)(1,0){6}{\sfr}
\multiput(.3,.3)(1,0){5}{\sfr}
\end{picture}}}

\newcommand{\ssixsixbox}
{\stln \lower2.6ex\hbox{
\begin{picture}(6.6,2.6)
\multiput(.3,1.3)(1,0){6}{\sfr}
\multiput(.3,.3)(1,0){6}{\sfr}
\end{picture}}}

\newcommand{\soneoneonebox}
{\stln \lower3.8ex\hbox{
\begin{picture}(1.6,3.6)
\multiput(.3,.3)(0,1){3}{\sfr}
\end{picture}}}

\newcommand{\stwooneonebox}
{\stln \lower3.8ex\hbox{
\begin{picture}(2.6,3.6)
\multiput(.3,.3)(0,1){3}{\sfr}
\put(1.3,2.3){\sfr}
\end{picture}}}

\newcommand{\stwotwoonebox}
{\stln \lower3.8ex\hbox{
\begin{picture}(2.6,3.6)
\multiput(.3,.3)(0,1){3}{\sfr}
\put(1.3,1.3){\sfr}
\put(1.3,2.3){\sfr}
\end{picture}}}

\newcommand{\stwotwotwobox}
{\stln \lower3.8ex\hbox{
\begin{picture}(2.6,3.6)
\multiput(.3,.3)(0,1){3}{\sfr}
\multiput(1.3,.3)(0,1){3}{\sfr}
\end{picture}}}

\newcommand{\sthreeoneonebox}
{\stln \lower3.8ex\hbox{
\begin{picture}(3.6,3.6)
\multiput(.3,.3)(0,1){3}{\sfr}
\put(1.3,2.3){\sfr}
\put(2.3,2.3){\sfr}
\end{picture}}}

\newcommand{\sthreetwoonebox}
{\stln \lower3.8ex\hbox{
\begin{picture}(3.6,3.6)
\multiput(.3,.3)(0,1){3}{\sfr}
\put(1.3,2.3){\sfr}
\put(2.3,2.3){\sfr}
\put(1.3,1.3){\sfr}
\end{picture}}}

\newcommand{\sthreetwotwobox}
{\stln \lower3.8ex\hbox{
\begin{picture}(3.6,3.6)
\multiput(.3,.3)(0,1){3}{\sfr}
\multiput(1.3,.3)(0,1){3}{\sfr}
\put(2.3,2.3){\sfr}
\end{picture}}}

\newcommand{\sthreethreeonebox}
{\stln \lower3.8ex\hbox{
\begin{picture}(3.6,3.6)
\multiput(.3,2.3)(1,0){3}{\sfr}
\multiput(.3,1.3)(1,0){3}{\sfr}
\put(.3,.3){\sfr}
\end{picture}}}

\newcommand{\sthreethreetwobox}
{\stln \lower3.8ex\hbox{
\begin{picture}(3.6,3.6)
\multiput(.3,2.3)(1,0){3}{\sfr}
\multiput(.3,1.3)(1,0){3}{\sfr}
\put(.3,.3){\sfr}
\put(1.3,.3){\sfr}
\end{picture}}}

\newcommand{\sthreethreethreebox}
{\stln \lower3.8ex\hbox{
\begin{picture}(3.6,3.6)
\multiput(.3,2.3)(1,0){3}{\sfr}
\multiput(.3,1.3)(1,0){3}{\sfr}
\multiput(.3,.3)(1,0){3}{\sfr}
\end{picture}}}

\newcommand{\sgencolbox}
{\stln \lower8.6ex\hbox{
\begin{picture}(1.6,7.6)
\multiput(.3,4.3)(0,1){3}{\sfr}
\put(.3,1.3){\framebox(1,3){$\vdots$}}
\put(.3,.3){\sfr}
\end{picture}}}

\newcommand{\sgenrowonebox}
{\stln \lower2.6ex\hbox{
\begin{picture}(7.6,2.6)
\multiput(.3,1.3)(1,0){3}{\sfr}
\put(3.3,1.3){\framebox(3,1){$\cdots$}}
\put(6.3,1.3){\sfr}
\put(.3,.3){\sfr}
\end{picture}}}

\newcommand{\sgenrowtwobox}
{\stln \lower2.6ex\hbox{
\begin{picture}(7.6,2.6)
\multiput(.3,1.3)(1,0){3}{\sfr}
\put(3.3,1.3){\framebox(3,1){$\cdots$}}

\put(6.3,1.3){\sfr}
\put(.3,.3){\sfr}
\put(1.3,.3){\sfr}
\end{picture}}}

\newcommand{\marcshapirowbox}
{\stln \lower2.6ex\hbox{
\begin{picture}(12.6,2.6)
\multiput(.3,1.3)(1,0){3}{\sfr}
\put(3.3,1.3){\framebox(3,1){$\cdots$}}
\put(6.3,1.3){\sfr}
\put(7.3,1.3){\sfr}
\put(8.3,1.3){\framebox(3,1){$\cdots$}}
\put(11.3,1.3){\sfr}
\put(7.3,1){$\underbrace{~~~~~~~~~~~~~~~}_{k}$}
\multiput(.3,.3)(1,0){3}{\sfr}
\put(3.3,.3){\framebox(3,1){$\cdots$}}
\put(6.3,.3){\sfr}
\put(.3,0){$\underbrace{~~~~~~~~~~~~~~~~~~~~~}_{n}$}
\end{picture}}}

\newcommand{\sgentworowonerowbox}
{\stln \lower2.6ex\hbox{
\begin{picture}(11.6,2.6)
\multiput(.3,1.3)(1,0){3}{\sfr}
\put(3.3,1.3){\framebox(3,1){$\cdots$}}
\put(6.3,1.3){\sfr}
\put(7.3,1.3){\framebox(3,1){$\cdots$}}
\put(10.3,1.3){\sfr}
\multiput(.3,.3)(1,0){3}{\sfr}
\put(3.3,.3){\framebox(3,1){$\cdots$}}
\put(6.3,.3){\sfr}
\end{picture}}}

\newcommand{\soneoneoneonebox}
{\stln \lower5ex\hbox{
\begin{picture}(1.6,4.6)
\multiput(.3,.3)(0,1){4}{\sfr}
\end{picture}}}

\newcommand{\soneoneoneoneoneonebox}
{\stln \lower7.4ex\hbox{
\begin{picture}(1.6,6.6)
\multiput(.3,.3)(0,1){6}{\sfr}
\end{picture}}}

\newcommand{\stwotwooneonebox}
{\stln \lower5ex\hbox{
\begin{picture}(2.6,4.6)
\multiput(.3,.3)(0,1){4}{\sfr}
\put(1.3,2.3){\sfr}
\put(1.3,3.3){\sfr}
\end{picture}}}

\newcommand{\sgenrowrowbox}
{\stln \lower2.6ex\hbox{
\begin{picture}(7.6,2.6)
\multiput(.3,1.3)(1,0){3}{\sfr}
\put(3.3,1.3){\framebox(3,1){$\cdots$}}
\put(6.3,1.3){\sfr}
\multiput(.3,.3)(1,0){3}{\sfr}
\put(3.3,.3){\framebox(3,1){$\cdots$}}
\put(6.3,.3){\sfr}
\end{picture}}}

\newcommand{\eq}{\begin{equation}}
\newcommand{\en}{\end{equation}}
\newcommand{\eqn}{\begin{eqnarray}}
\newcommand{\enn}{\end{eqnarray}}

\newcommand{\beq}{\begin{equation}}
\newcommand{\eeq}{\end{equation}}
\begin{document}
\begin{titlepage}
\begin{flushright}
  PSU-TH-201 \\
\end{flushright}
\begin{center}
{\bf 4D DOUBLETON CONFORMAL THEORIES, $CPT$ AND
II B STRING ON $AdS_{5} \times S^{5}$ } \\
\vspace{1cm}
{\bf M. G\"{u}naydin\footnote{Work supported in part by the
National Science Foundation under Grant Number PHY-9631332. \newline
e-mail: murat@phys.psu.edu}
, D. Minic\footnote{
e-mail: minic@phys.psu.edu}
and M. Zagermann} \footnote{
e-mail: zagerman@phys.psu.edu}  \\
\vspace{.5cm}
Physics Department \\
Penn State University\\
University Park, PA 16802 \\
\vspace{1cm}
{\bf Abstract}
\end{center}
We study  the unitary supermultiplets of the ${\cal{N}}=8$ $d=5$ 
anti-de Sitter (AdS)
superalgebra $SU(2,2|4)$ which is the symmetry algebra of the IIB string theory
on $AdS_5 \times S^5$. We give a complete classification of the doubleton
supermultiplets of $SU(2,2|4)$ which do not have 
a Poincare limit and correspond
to $d=4$ conformal field theories (CFT) living on the boundary of $AdS_5$.
The  CPT self-conjugate irreducible 
doubleton supermultiplet corresponds
to $d=4$ ${\cal{N}}=4$ super Yang-Mills theory. The other irreducible
doubleton supermultiplets come in CPT conjugate pairs. The maximum spin range
of the general doubleton supermultiplets is 2. In particular, there exists a
CPT conjugate 
pair of doubleton supermultiplets corresponding to the fields  of 
${\cal{N}}=4$ conformal supergravity in $d=4$ which can be coupled to 
${\cal{N}}=4$ super Yang-Mills theory in $d=4$. We also study the
"massless" supermultiplets of $SU(2,2|4)$ which can be obtained by 
tensoring two doubleton supermultiplets. The CPT self-conjugate
"massless" supermultiplet is the ${\cal{N}}=8$ graviton supermultiplet
in $AdS_5$. The other "massless" supermultiplets generally come in  conjugate
pairs and can have maximum spin range of 4. We discuss the implications of
our results for the conjectured CFT/AdS dualities.

\end{titlepage}

\renewcommand{\theequation}{\arabic{section} - \arabic{equation}}
\section{Introduction}
\setcounter{equation}{0}

Since the original conjecture of Maldacena \cite{mald}
relating the large N limits of certain conformal field theories
(CFT) in d-dimensions to M-theory/string theory
compactified to d+1-dimensional AdS spacetimes, a lot
of work has been done on CFT/AdS duality. Maldacena's
conjecture was originally based on 
certain  properties of the physics of N Dp-branes in the
near horizon limit \cite{ads} and the old knowledge about
10-d IIB supergravity
compactified on $AdS_5 \times S^5$, and
11-d supergravity compactified on $AdS_7 \times S^4$ and
$AdS_4 \times S^7$\cite{dfhn, mgnm, krv, ptn, mgnw, gnw, ss}.
In particular, Maldacena's conjecture was made
more precise in \cite{pol,witt}.

The relation between Maldacena's
conjecture and the dynamics of the singleton and doubleton fields that
live on the boundary of AdS 
spacetimes was reviewed in \cite{sfcf} and \cite{mgdm}.

In this paper we want to focus on the prime example of this
CFT/AdS duality, namely the duality between the large N limit
of ${\cal{N}}=4$ $SU(N)$ super Yang-Mills theory in $d=4$ and the IIB string
theory over $AdS_5 \times S^5$.

The ${\cal{N}}=4$ super Yang-Mills multiplet corresponds to the 
CPT self-conjugate irreducible doubleton supermultiplet of the ${\cal{N}}=8$
AdS superalgebra $SU(2,2|4)$ in $d=5$. There exist other doubleton 
supermultiplets of $SU(2,2|4)$ which are not CPT
self-conjugate. One of the goals of this paper is to
emphasize this point and give a complete list
of all doubleton supermultiplets of $SU(2,2|4)$. In particular, we find 
doubleton supermultiplets corresponding to ${\cal{N}}=4$ conformal
supergravity living on the boundary of $AdS_5$, which can be identified
with the $d=4$ Minkowski space.

The $d=4$, ${\cal{N}}=4$ super Yang-Mills matter can be coupled to 
${\cal{N}}=4$ conformal supergravity. Since the resulting $d=4$
theory is conformally invariant we expect that  
Maldacena's conjecture can be generalized so as to
 include the degrees of freedom of the
conformal supergravity sector.

We also study "massless" supermultiplets of 
${\cal{N}}=8$ $AdS_5$ superalgebra $SU(2,2|4)$. The long
"massless" supermultiplets have spin range 4. However, there 
exist other "massless" supermultiplets whose spin range is less than 4.

The paper is organized as follows: In section 2. we present a short
review of the oscillator method. Section 3. gives the general construction
of the positive energy representations of $SU(2,2)$. In section 4. we
give a complete list of doubleton representations of $SU(2,2)$.
Likewise, in sections 5. and 6. we give complete lists of 
"massless" and "massive" positive energy representations of 
$SU(2,2)$. In section 7. the oscillator method for the general supergroup
$SU(m,n|p+q)$ is reviewed. In section 8. we give a complete list of
doubleton supermultiplets of $SU(2,2|4)$ and in section 9.
we study the "massless" irreducible supermultiplets of 
$SU(2,2|4)$.
We conclude the article with a discussions of the implications our results
have for CFT/AdS duality.

\section{Short Review of the Oscillator Method}
\setcounter{equation}{0}

In \cite{mgcs} a general oscillator method was developed for
constructing the unitary irreducible representations (UIR) of
the lowest (or highest) weight type of non-compact groups.
The oscillator method yields the UIR's of lowest weight type of a 
noncompact group over the Fock space of a set of bosonic 
oscillators. To achieve this one realizes the generators of the 
noncompact group as bilinears of  sets of bosonic oscillators
transforming  in a finite dimensional representation of its maximal
compact subgroup. The minimal realization of these generators requires
either one or two sets of bosonic annihilation and 
creation operators transforming
 irreducibly under its maximal compact subgroup. These minimal 
 representations are 
fundamental in that all the other ones can be obtained from the minimal
representations by a simple tensoring procedure.  
 These fundamental representations
are nothing but a generalization of the celebrated remarkable representations
of the $AdS_4$ group $SO(3,2)$ discovered by Dirac
\cite{pam}
long time ago, which
were later named singletons \cite{fron} (indicating
the fact that the remarkable representations of Dirac 
corresponding to the fields living on the boundary of
$AdS_4$ are singular when the Poincare limit is taken).
In the language of the oscillator method, these singleton representations
require a single set of oscillators transforming in the fundamental
representation of the maximal compact sugroup of the covering group
$Sp(4,R)$ of $SO(3,2)$ \cite{mgnw, mg81, mg} (a fact that meshes nicely
with the name singleton).
In some cases (as with the $AdS_5$ group
$SU(2,2)$) the fundamental representations require two sets of oscillators,
and they were called doubletons in \cite{gnw, mgnm}. The general oscillator 
construction of the lowest (or highest) weight representations 
of non-compact supergroups (i.e. the case when the even subgroup
is non-compact) was given in \cite{ibmg}. 
 The oscillator method was further developed and applied to the spectra
of Kaluza-Klein supergravity theories in references \cite{mgnm, mgnw, gnw}.

A non-compact group $G$ that admits unitary representations of the
lowest weight type has a maximal compact subgroup $G^{0}$  of the form
$G^{0} = H \times U(1)$ with respect to whose Lie algebra $g^0$  one has
a three grading of the Lie algebra $g$ of $G$,
\eq
g = g^{-1} \oplus g^{0} \oplus g^{+1}
\en
which simply means that the commutators of elements of grade
$k$ and $l$ satisfy
\eq
[g^{k},g^{l}] \subseteq g^{k+l}.
\en
Here $g^{k+l} = 0$ for $|k+l| >1$.

For example, for $SU(1,1)$ this corresponds to the standard
decomposition
\eq
g = L_{+} \oplus L_{0} \oplus L_{-}
\en
where
\eqn
[L_{0},L_{\pm}] &= &\pm L_{\pm} \cr
[L_{+},L_{-}] &=&2L_{0}
\enn

The three grading is determined by the generator $E$ of the
$U(1)$ factor of the maximal compact subgroup
\eqn
[E,g^{+1}] &=&g^{+1} \cr
[E,g^{-1}] &=&-g^{-1} \cr
[E,g^{0}] &=& 0
\enn

In most physical applications $E$  turns out
to be the energy operator.
In such cases the unitary lowest weight representations
correspond to positive energy representations.

The  bosonic annihilation
and creation operators in terms of which one realizes the generators
of $G$ transform typically in the fundamental
and its conjugate representation of $H$.
In the Fock space $\cal{H}$ of all the oscillators one
chooses a set of  states
$|\Omega \rangle$ which transform irreducibly under $H \times U(1)$
and are annihilated by all the generators in $g^{-1}$.
Then by acting on $|\Omega \rangle$ with generators in
$g^{+1}$ one obtains an infinite set of states
\eq
|\Omega \rangle ,\quad  g^{+1}|\Omega \rangle ,\quad
g^{+1} g^{+1}|\Omega \rangle , ...
\en
which form an UIR of the lowest weight (positive energy) type
of $G$. The infinite set of states thus obtained corresponds to the
decomposition of the UIR of $G$ with respect to its maximal
compact subgroup.

As we have already emphasized, whenever
 we can realize the generators of $G$ in
terms of a single set of bosonic creation (and annihilation )
operators transforming in an irreducible representation (and its conjugate) of
the compact subgroup $H$ then the corresponding
UIRs  will be called singleton representations
and  there exist two such representations for a given group
 $G$. For the AdS group in $d=4$  the singleton
representations correspond to scalar and spinor fields .
In certain cases  we need  a minimum of two sets of 
bosonic creation  and annihilation 
operators transforming irreducibly under $H$ to realize the
generators of $G$. In such cases  the
corresponding UIRs  are called doubleton
representations and there exist
infinitely many doubleton representations
of $G$ corresponding to fields of different
"spins".
For example, the non-compact
group $Sp(2N,R)$ with the maximal compact subgroup $U(N)$ 
admits singleton representations \cite{mgnw, mgcs, mgsh}.
On the other hand, the non-compact groups $SO^{*}(2N)$ \cite{gnw, mgcs} and
$SU(N,M)$ \cite{mgcs,ibmg} with maximal compact subgroups $U(N)$
and $S(U(M)\times U(N))$ respectively, admit doubleton representations.

 The noncompact supergroups also admit either singleton or doubleton
supermultiplets corresponding to some minimal fundamental unitary irreducible
representations, in terms of which one can contruct all the other
UIR's of the lowest weight type by a simple tensoring procedure.
For example, the non-compact supergroup $OSp(2N/2M,R)$ with the even subgroup
$SO(2N)\times Sp(2M,R)$  admits singleton
supermultiplets, while $OSp(2N^*|2M)$ and $SU(N,M|P)$ with even subgroups
$SO^*(2N) \times USp(2M)$ and $SU(N,M)\times SU(P) \times U(1)$ admit
doubleton supermultiplets.

Even though the Poincare limit of the singleton (or doubleton) representations
is singular, the tensor product of two singleton (or doubleton)
representations decomposes into an infinite set of "massless"
irreducible representations which do have a smooth Poincare limit
\cite{mgnw, fron, mg81}. Based on this fact the following
definition of "massless" representations in AdS space-time was
proposed in \cite{mg}:

{\it A representation (or a supermultiplet) of an AdS group (or supergroup)
is "massless" if it occurs in the decomposition of the
tensor product of two singleton or two doubleton representations
(or supermultiplets).}

This should be taken as a working definition which agrees with
some other definitions of "masslessness" in $d \leq 7$.
Note also that recent work on CFT/AdS duality gives support to the above
definition from a dynamical point of view \cite{witt, ads2}.
Furthermore, tensoring more than two singletons or doubletons
representations leads to "massive" representations of
AdS groups and supergroups \cite{mg}.

\section{Oscillator Construction of the Positive Energy Representations
of $SU(2,2)$}
\setcounter{equation}{0}

Unitary representations of the covering group $SU(2,2)$ of the conformal
group $SO(4,2)$ in $d=4$ have been studied extensively \cite{frad}.
The group $SO(4,2)$ is also the AdS group of $d=5$ spacetime with
Lorentzian signature.
Positive energy or equivalently the lowest weight representations of
$SU(2,2)$ can be constructed very simply by the oscillator method outlined
in the previous section \cite{mgnm, ibmg}.

Let us denote the two $SU(2)$ subgroups of $SU(2,2)$ as $SU(2)_L$ and
$SU(2)_R$ respectively. The generator of the Abelian factor in the maximal
compact sugroup of $SU(2,2)$ is the AdS energy operator in $d=5$
(or the conformal Hamiltonian in $d=4$ whose eigenvalues give the
conformal dimensions) and will be denoted as E.
To construct the positive $AdS_5$ energy (or d=4 conformal) representations
we realize the generators of $SU(2,2)$ as bilinears of pairs of
bosonic oscillators transforming in the fundamental
representation of the two $SU(2)$ subgroups. The oscillators
satisfy the canonical commutation relations
\eq
[a_i(\xi), a^j(\eta)] = \delta_{i}^{j} \delta_{\xi \eta} \quad
i,j =1,2.
\en
\eq
[b_r(\xi), b^s(\eta)] = \delta_{r}^{s} \delta_{\xi \eta} \quad
r,s=1,2
\en
Here $\xi, \eta = 1,...,P$ label different generations of
oscillators and
\eq
[a_i(\xi), b_r(\eta)] = [a_i(\xi), b^r(\eta)] = [a_i(\xi), a_j(\eta)] = [b_r(\xi), b_s(\eta)]=0
\en

The bosonic oscillators with an upper index ($i$ or $r$)
are creation operators while those with lower indices are
annihilation operators. The vacuum vector is defined as
\eq
a_i(\xi) |0 \rangle = 0 = b_r(\xi) |0 \rangle
\en
for all values of $i,r,\xi$.
The non-compact generators of $SU(2,2)$ are realized by the following
bilinears
\eq
L_{ir}= {\vec{a}}_i \cdot {\vec{b}}_{r} \quad
L^{ir}= {\vec{a}}^{i} \cdot {\vec{b}}^{r}
\en
where ${\vec{a}}_{i} \cdot {\vec{b}}_{r} = \sum_{\xi=1}^{P} a_i(\xi) b_r(\xi)$
etc.
They close into the generators of the compact subgroup
$SU(2)_L \times SU(2)_R \times U(1)$
\eqn
[L_{ir}, L^{js}] &=& \delta_{r}^{s} L^{j}_{i} +
\delta_{i}^{j} R^{s}_{r} +\delta_{i}^{j} \delta_{r}^{s} E \cr
[L_{ir}, L_{js}] &=&[L^{ir}, L^{js}] =0
\enn
where
\eqn
L^{j}_{i}&=& {\vec{a}}^{j} \cdot {\vec{a}}_{i} -
{1 \over 2} \delta_{i}^{j}{\vec{a}}^{l} \cdot {\vec{a}}_{l} \cr
R^{r}_{s}&=& {\vec{b}}^{r} \cdot \vec{b_s} -
{1 \over 2} \delta_{s}^{r}\vec{b^t} \cdot \vec{b_t} \cr
E&=& {1 \over 2} ({\vec{a}}_{i} \cdot {\vec{a}}^{i} +
{\vec{b}}^{r} \cdot {\vec{b}}_{r})
\enn
Here $L^{j}_{i}$ and $R^{r}_{s}$ are the generators of
$SU(2)_L$ and $SU(2)_R$, respectively.

Defining the number operators
\eqn
N_a &=& {\vec{a}}^{i} \cdot {\vec{a}}_{i} = \sum_{i=1}^{2}
\sum_{\xi =1}^{P} a^i(\xi) a_i(\xi) \cr
N_b &=& {\vec{b}}^{r} \cdot {\vec{b}}_{r} \cr
 N &=& N_a + N_b
\enn
we can write the AdS energy operator E as \footnote{ From this expression
it is clear that the eigenvalues of the energy operator in $AdS_5$ (or 
conformal dimensions in d=4) take on integer or half-integer values.
To get the generic, real, values of the conformal dimension (which
includes the anomalous dimension) one needs to take the infinite
covering of the conformal group $SU(2,2)$ and consider its lowest weight
UIRs \cite{frad}. }
\eq
E = {1 \over 2} (N_a +N_b + 2P) = {1 \over 2} N + P
\en

The quadratic Casimir operator $C_2$ of $SU(2,2)$ is uniquely defined
up to an
overall multiplicative constant. We choose this constant such
that
\eq
C_2 = -{1 \over 2}( L_{ir} L^{ir} +  L^{ir} L_{ir})
+ {1 \over 2} ( L^{i}_{j} L^{j}_{i} + R^{r}_{s} R^{s}_{r} + E^2)
\en
This expression can be rewritten in terms of number operators
$N_a, N_b$ and $N$ for the case of $P=1$
\eq
C_2 = ({N \over 2} + 1) ({N \over 2} -3) +  N_a ({N_a \over 2} + 1)
+  N_b ({N_b \over 2} + 1)
\en

The positive energy irreducible unitary representations
of $SU(2,2)$ are uniquely defined by a lowest weight vector
$| \Omega \rangle$ transforming irreducibly under the maximal compact subgroup
$S(U(2) \times U(2))$ and that is annihilated by
$L_{ir}$
\eq
L_{ir} | \Omega \rangle =0
\en

Then by acting on $| \Omega \rangle$ repeatedly with the generators
$L^{ir}$ one generates an infinite set of states 
\eq
|\Omega \rangle ,\quad  L^{ir}|\Omega \rangle ,\quad
 L^{ir} L^{js}|\Omega \rangle , ...
\en
that form the basis of a unitary irreducible representation
of $SU(2,2)$ \cite{mgnm, ibmg}.
This infinite set of states corresponds to the decomposition of a
positive energy UIR of $SU(2,2)$ with respect to the maximal
compact subgroup. They can be
identified with the Fourier modes of a field in $AdS_5$ that is
uniquely defined by the lowest weight vector. If the lowest weight
vector transforms in the $(j_{L}, j_{R})$ representation of
$SU(2)_L \times SU(2)_R$ and has AdS energy
$E$ the corresponding field in
$AdS_5$ will be denoted as
\eq
\Xi_{(j_{L},j_{R})}(E)
\en
Note that the local fields that transform covariantly under
the Lorentz group in $d=5$ correspond in general to direct sums 
of such fields and their conjugates.

The eigenvalues of
the quadratic Casimir operator $C_2$ of $SU(2,2)$  can be, 
in general (i.e. for arbitrary $P$), expressed in the following form
\eq
C_{2} = E (E -4) +2j_{L}(j_{L}+1) + 2j_{R}(j_{R}+1)
\en
where $E$ denotes the AdS energy and $j_{L}, j_{R}$
the two $SU(2)$ quantum numbers.
For $P=1$, $j_{L}, j_{R}$ are determined by the number operators
$N_{a},  N_{b}$
 as follows
\eq
j_{L} = {N_{a} \over 2}, \quad j_{R} = {N_{b} \over 2}
\en

Similarly, the eigenvalues of the  cubic, $C_{3}$,  and  quartic Casimir
operators,$C_{4}$, can be expressed in terms of  $E, j_{L}$ and $j_{R}$ as
\eq
C_{3} = - (E-2) [j_{L}(j_{L} + 1) - j_{R}(j_{R} + 1)]
\en
\eqn
C_{4} &= {1 \over 4} (E -2)^{4} - (E-2)^{2}
[j_{L}(j_{L} + 1) + j_{R}(j_{R} + 1) +1] \cr
 & + 4 j_{L}(j_{L} + 1)j_{R}(j_{R} + 1)
\enn

\section{Doubleton Representations of the $AdS_5$ Group $SU(2,2)$}
\setcounter{equation}{0}

The minimal oscillator realization of $SU(2,2)$ requires a
single pair of oscillators i.e. $P=1$, corresponding
to doubleton representations \cite{mgnm} that have no Poincare limit.
Possible lowest weight vectors in this case are of the form
\eq
a^{i_1}...a^{i_{n_L}}|0\rangle
\en
and 
\eq
b^{r_1}...b^{r_{n_R}}|0\rangle
\en
where $n_L$ and $n_R$
are some non-negative integers (including zero).

The $AdS_5$ fields corresponding to the UIR's defined by these lowest weight
vectors are
\eq
a^{i_1}...a^{i_{n_L}}|0\rangle
\Leftrightarrow \Xi_{(\frac{n_L}{2},0)}({n_L \over 2} +1)
\en
\eq
b^{r_1}...b^{r_{n_R}}|0\rangle
\Leftrightarrow \Xi_{(0,{n_R \over 2})}({n_R \over 2} +1)
\en

The eigenvalues of the quadratic Casimir operator $C_2$
on these $AdS_5$ fields are given by
\eq
C_2 \Xi_{({n_L \over 2},0)}({n_L \over 2} +1) 
= 3 ({n_L \over 4}^2 -1) \Xi_{({n_L \over 2},0)}({n_L \over 2} +1)
\en
\eq
C_2 \Xi_{(0,{n_R \over 2})}({n_R \over 2} +1) 
= 3 ({n_R \over 4}^2 -1) \Xi_{(0,{n_R \over 2})}({n_R \over 2} +1)
\en

We should note that since the Poincare limit of the 
doubleton representations of the $AdS_5$
group is singular they are to be interpreted as 
fields living on the boundary of
$AdS_5$ space which can be identified with the 
four dimensional Minkowski space-time
with some points added \cite{witt}. 
Then the group $SU(2,2)$ acts as the conformal
group on the boundary.  

\section{"Massless" Representations of the $AdS_5$ Group $SU(2,2)$}
\setcounter{equation}{0}

"Massless" representations of the $AdS_5$ group $SU(2,2)$
are obtained when we
take two pairs ($P=2$) of oscillators  in the oscillator construction.
In this case we have   lowest weight vectors with the same 
$SU(2)_L \times SU(2)_R$ transformation 
properties as in the case of doubletons,
but with $AdS$ energies one unit higher :
\eq
a^{(i_1}...a^{i_{n_L})} |0\rangle \Leftrightarrow
\Xi_{({n_L \over 2},0)}({n_L \over 2} +2)
\en
\eq
b^{(r_1}...b^{r_{n_R})}|0\rangle \Leftrightarrow
\Xi_{(0,{n_R \over 2})}({n_R \over 2} +2)
\en
where the round brackets indicate symmetrization of indices.

We also have a new type of lowest weight vectors with both $j_{L}$ and $j_{R}$
nonvanishing, which have no analogs in the case of doubletons:
\eq
a^{i_1}(1)...a^{i_{n_L}}(1) b^{r_1}(2)...b^{r_{n_R}}(2)
|0\rangle \Leftrightarrow
\Xi_{({n_L \over 2}, {n_R \over 2})}({{n_L + n_R} \over 2} +2)
\en
The eigenvalues of the quadratic Casimir operator on the UIR's 
defined by the above lowest weight vectors are
\eq
C_2 \Xi_{({n_L \over 2},0)}({n_L \over 2} +2) 
= ({3 \over 4} n_{L}^2 +n_{L} -4)
  \Xi_{({n_L \over 2},0)}({n_L \over 2} +2)
\en
\eq
C_2 \Xi_{(0,{n_R \over 2})}({n_R \over 2} +2) 
= ({3 \over 4} n_{R}^2 +n_{R} -4) \Xi_{(0,{n_R \over 2})}({n_R \over 2} +2)
\en
\eq
C_2 \Xi_{({n_L \over 2},{n_R \over 2})}(\frac{n_L +n_R}{ 2} +2) 
= ({3 \over 4} (n_{R}^2 +n_{L}^{2}) +n_{R}  
 +n_{L} + { {n_{L} n_{R}} \over 2} -4) 
 \Xi_{({n_L \over 2},{n_R \over 2})}
({{n_R + n_L} \over 2} +2)
\en

In addition to the above lowest weight vectors for $P=2$
we have the following possible lowest weight vectors and the
corresponding fields:
\eq
a^{[j_{1}}a^{k_{1}]} a^{[j_{2}}a^{k_{2}]}...a^{[j_{L}}a^{k_{L}]}
a^{(i_{1}}...a^{i_{n_{L}})} |0 \rangle 
\Leftrightarrow \Xi_{({{n_{L}} \over 2},0)}({n_{L} \over 2} +L +2) 
\en
\eq
b^{[s_{1}}b^{t_{1}]} b^{[s_{2}}b^{t_{2}]}...b^{[s_{R}}b^{t_{R}]}
b^{(r_{1}}...b^{r_{n_{R}})} |0 \rangle
\Leftrightarrow \Xi_{(0,{{n_{R}} \over 2})} ({n_{R} \over 2} +R+2)
\en
where the square brackets indicate antisymmetrization of indices.

\section{"Massive" Representations of the $AdS_5$ Group 
$SU(2,2)$}
\setcounter{equation}{0}

"Massive" representations of the $AdS_5$ group $SU(2,2)$ can be
obtained by the oscillator method by taking $P >2$. For example,
the lowest weight vectors (5.1), (5.2) and (5.3) correspond to
"massive" fields for $P >2$:
\eq
a^{(i_{1}}...a^{i_{n_{L}})} |0 \rangle
\Leftrightarrow \Xi_{({n_{L} \over 2},0)} ({n_{L} \over 2} + P)
\en
\eq
b^{(r_{1}}...b^{r_{n_{R}})} |0\rangle
\Leftrightarrow \Xi_{(0,{n_{R} \over 2})} ({n_{R} \over 2} + P)
\en
\eq
a^{(i_{1}}(1)...a^{i_{n_{L}})}(1) 
b^{(r_{1}}(2)...b^{r_{n_{R}})}(2) |0\rangle
\Leftrightarrow \Xi_{({n_{L} \over 2}, {n_{R} \over 2})} ({{n_{L}+n_{R}} 
\over 2} + P)
\en
Some of these "massive" fields with "spin" less than or equal to two 
appear in the spectrum of the $S^5$ compactification 
of IIB theory \cite{mgnm,krv}.
We should note, however, that the mass operator is not an invariant operator
of the $AdS_5$ group $SU(2,2)$.   
To illustrate the problems associated with
defining the concept of mass in $AdS$ space, consider the
UIR associated with the vacuum $|0\rangle$ chosen as the
lowest weight vector (lwv) for $P=L+2$
\eq
|0\rangle \Leftrightarrow \Xi_{(0,0)} (L+2)
\en

On the other hand, the lowest weight vector
\eq
a^{[i_{1}}a^{j_{1}]}...a^{[i_{L}} a^{j_{L}]} |0\rangle
\Leftrightarrow \Xi_{(0,0)} (L+2)
\en
with $P=2$ yields the same UIR.
The analysis based on the wave equations in $AdS$ spacetimes suggests
that the UIR defined by $|0\rangle$ for $P=L+2$ $(L>0)$ is "massive"
\cite{krv}. On the other hand, the state
$a^{[i_{1}}a^{j_{1}]}...a^{[i_{L}} a^{j_{L}]} |0\rangle
$
for $P=2$
occurs in the tensor product of two doubletons and must be "massless"
\cite{mgnm, fron}.

The resolution of the puzzle is that the standard analysis using
wave equations assumes that the field in question is an elementary
field \cite{krv}.  An elementary scalar $\Phi_{(0,0)} (L+2)$ field
corresponding to the lowest weight vector  $|0\rangle$ with $P=L+2$ has
the same $AdS$ energy as the composite scalar built out of 
$2L$ massless "spin 1/2" fields $\Psi$ of the form
$(\bar{\Psi} \Psi) ... (\bar{\Psi} \Psi)$.

Much more useful and unambiguous concept than mass is that of $AdS$ energy or
conformal dimension (conformal energy) when
$SU(2,2)$ is interpreted as the 4-d conformal group \cite{witt}. 
When we consider $SU(2,2)$  as the conformal group then the
conformal dimension of a composite operator is simply given
by the sum of the conformal dimensions of its elementary 
constituents and we do not have any interpretational problems.

\section{$SU(m,n|p+q)$ via the Oscillator Method }
\setcounter{equation}{0}

The symmetry group of the
compactification of  type IIB superstring over the five sphere
is the supergroup $SU(2,2|4)$ with the even subgroup
$SU(2,2) \times SU(4) \times U(1)$ where $SU(4)$ is
the isometry group of the five sphere \cite{mgnm}. The Abelian $U(1)$
factor comes directly from the ten dimensional theory itself
and is the subgroup of the $SU(1,1)$ symmmetry of the IIB
supergravity. This $U(1)$ generator commutes with all the other
generators and acts like a central charge. Therefore, $SU(2,2|4)$
is not a simple Lie superalgebra. By factoring out this Abelian
 ideal one obtains a simple Lie superalgebra, denoted as $PSU(2,2|4)$,
whose even subalgebra is simply $SU(2,2)\times SU(4)$
\footnote{ In \cite{mgnm} the symmetry supergroup of the $S^5$ 
compactification of IIB theory was denoted as $U(2,2|4)$ to stress
the fact that it contains an Abelian ideal.}. By orbifolding the five
sphere with some discrete subgroup one can obtain consistent backgrounds
for the compactification of IIB superstring \cite{ads1}. These
backgrounds have fewer supersymmetries corresponding to supergroups
of the form $SU(2,2|k)$ with $k<4$. 
In the following sections we apply the general 
oscillator method outlined in the
introduction to construct positive energy unitary irreducible 
representations of
$SU(2,2|4)$. Before specializing to the 
supergroup $SU(2,2|4)$ relevant for the IIB
superstring we shall discuss the oscillator realization of 
general supergroups of the form $SU(m,n|p+q)$ following \cite{ibmg,mgnm}.

The superalgebra $SU(m,n|p+q)$ has a three graded decomposition with
respect to its compact subsuperalgebra $SU(m|p)\times SU(n|q) \times U(1)$
\eq
g = L^{+} \oplus L^{0} \oplus L^{-}
\en
where
\eqn
[L^{0},L^{\pm}] &= & L^{\pm} \cr
[L^{+},L^{-}] &=&L^{0} \cr
[L^{+},L^{+}] &=& 0=[L^{-},L^{-}]
\enn

Here $L^{0}$ represents the generators of
$SU(m|p) \times SU(n|q) \times U(1)$.

The Lie superalgebra $SU(m,n|p+q)$
 can be realized in terms of bilinear combinations of bosonic and
fermionic annihilation and creation operators $\xi_{A}$
($\xi^{A}={\xi_{A}}^{\dagger}$) and $\eta_{M}$
($\eta^{M}={\eta_{M}}^{\dagger}$)
which transform covariantly and contravariantly
under the  $SU(m|p)$ and $SU(n|q)$ subsupergroups of $SU(m,n|p+q)$
\eq
\xi_{A} = \left(\matrix{a_{i} \cr
                        \alpha_{\mu} \cr} \right) ,\quad
\xi^{A} = \left(\matrix{a^{i} \cr
                        \alpha^{\mu} \cr} \right)
\en
and
\eq
\eta_{M} = \left(\matrix{b_{r} \cr
                        \beta_{x} \cr} \right) , \quad
\eta^{M} = \left(\matrix{b^{r} \cr
                        \beta^{x} \cr} \right)
\en
with $i,j=1,2, .., m$; $\mu,\nu=1,2,..,p$; $r,s=1,2,..,n$; $x,y=1,2,..,q$ and
\eq
[a_i, a^j] = \delta_{i}^{j} , \quad
\{\alpha_{\mu}, \alpha^{\nu}\} = \delta_{\mu}^{\nu}
\en
\eq
[b_r, b^s] = \delta_{r}^{s} , \quad
\{\beta_{x}, \beta^{y}\} = \delta_{x}^{y}
\en
Again we denote the annihilation and creation operators with lower and
upper indices, respectively. 
The generators of $SU(m,n|p+q)$ are given in terms of the above 
superoscillators
as
\eqn
L^{-} &=& {\vec{\xi}}_{A} \cdot {\vec{\eta}}_{M} \cr
L^{0} &=& {\vec{\xi}}^{A} \cdot {\vec{\xi}}_{B}
\oplus {\vec{\eta}}^{M} \cdot {\vec{\eta}}_{N} \cr
L^{+} &=& {\vec{\xi}}^{A} \cdot {\vec{\eta}}^{M}
\enn
where the arrows over $\xi$ and $\eta$ indicate that we are taking an
arbitrary number $P$ of 
superoscillators and the dot represents the summation
over the internal index $k = 1,...,P$, i.e
${\vec{\xi}}_{A} \cdot {\vec{\eta}}_{M} \equiv \sum_{k=1}^{P}
{\xi_{A}}(k){\eta_{M}}(k)$.

The $SU(p+q)$ generators, written in terms of fermionic
oscillators $\alpha$ and $\beta$, read as follows
\eqn
A^{\nu}_{\mu} &=& {\vec{\alpha}}^{\nu} \cdot {\vec{\alpha}}_{\mu}
-{1 \over p} \delta^{\nu}_{\mu} N_{\alpha} \cr
B^{y}_{x} &=& {\vec{\beta}}^{y} \cdot {\vec{\beta}}_{x}
-{1 \over q} \delta^{y}_{x} N_{\beta} \cr
C &=& -{1 \over p} N_{\alpha} - {1 \over q} N_{\beta} +P \cr
L_{\mu x} &=& {\vec{\alpha}}_{\mu} \cdot {\vec{\beta}}_{x},
\quad L^{x \mu} = {\vec{\beta}}^{x} \cdot {\vec{\alpha}}^{\mu}
\enn
where $N_{\alpha}={\vec{\alpha}}^{\nu} \cdot {\vec{\alpha}}_{\nu} $
and $N_{\beta} = \vec{\beta^{x}} \cdot {\vec{\beta}}_{x}$
are the fermionic number operators.

Similarly, the $SU(m,n)$ generators, written in terms of
bosonic oscillators $a$ and $b$, read 
\eqn
L_{ir} &=& {\vec{a}}_{i} \cdot \vec{b_{r}}, \quad
L^{ir} = {\vec{a}}^{i} \cdot \vec{b^{r}} \cr
L^{k}_{i} &=& {\vec{a}}^{k} \cdot {\vec{a}}_{i}
-{1 \over m} \delta^{k}_{i}N_{a} \cr
R^{r}_{s} &=& \vec{b^{r}} \cdot \vec{b_{s}}
-{1 \over n} \delta^{r}_{s} N_{b}\cr
E &=& {1 \over m} N_a + {1 \over n} N_b + P
\enn
where $N_{a} \equiv {\vec{a}}^{i} \cdot {\vec{a}}_{i}, 
N_{b} \equiv {\vec{b}}^{r} \cdot {\vec{b}}_{r}$ are the
bosonic number operators.

The following closure relations are valid
\eqn
[{\vec{a}}_{i} \cdot {\vec{b}}_{r},{\vec{b}}^{t} \cdot {\vec{a}}^{k}]
&=& \delta^{t}_{r} L^{k}_{i} + \delta^{k}_{i} R^{t}_{r}
+ \delta^{k}_{i}\delta^{t}_{r} E \cr
[{\vec{\alpha}}_{\mu} \cdot {\vec{\beta}}_{x},
{\vec{\beta}}^{z} \cdot {\vec{\alpha}}^{\rho}]
&=& -\delta^{z}_{x} A^{\rho}_{\mu} - \delta^{\rho}_{\mu} B^{z}_{x}
+ \delta^{z}_{x} \delta^{\rho}_{\mu} C \cr
\{ {\vec{a}}_{i} \cdot {\vec{\beta}}_{x},{\vec{\beta}}^{z}
\cdot {\vec{a}}^{k} \}
&=& \delta^{z}_{x} L^{k}_{i} - \delta^{k}_{i} B^{z}_{x}
+ {1 \over 2}\delta^{z}_{x}\delta^{k}_{i} {(E+C+D-F)} \cr
\{ {\vec{\alpha}}_{\mu} \cdot {\vec{b}}_{r},
{\vec{b}}^{t} \cdot {\vec{\alpha}}^{\rho} \}
&=& -\delta^{t}_{r} A^{\rho}_{\mu} + \delta^{\rho}_{\mu} R^{t}_{r}
+ {1 \over 2}\delta^{t}_{r} \delta^{\rho}_{\mu} {(E+C-D+F)} \cr
\{ {\vec{a}}^{i} \cdot {\vec{\alpha}}_{\mu},{\vec{\alpha}}^{\rho}
\cdot {\vec{a}}_{k}\}
&=& \delta^{\rho}_{\mu} L^{i}_{k} + \delta^{i}_{k} A^{\rho}_{\mu}
+ {1 \over 2}\delta^{\rho}_{\mu} \delta^{i}_{k} {(E-C+D-F)} \cr
\{ {\vec{b}}^{r} \cdot {\vec{\beta}}_{x},{\vec{\beta}}^{z}
\cdot {\vec{b}}_{t} \}
&=& \delta^{z}_{x} R^{r}_{t} + \delta^{r}_{t} B^{z}_{x}
+ {1 \over 2}\delta^{z}_{x}\delta^{r}_{t} {(E-C-D+F)}
\enn
where $D$ and $F$ are defined as
\eqn
D &=& {1 \over m} N_a - {1 \over q} N_{\beta} + P  \cr
F &=& {1 \over n} N_b - {1 \over p} N_{\alpha} + P
\enn
Note that only three of the four $U(1)$ charges $E,C,D,F$ are 
linearly independent. In what follows, we will choose $E, C$ and $(D-F)$
as the linearly independent $U(1)$ generators.

The quadratic Casimir operator must be of the form
\eqn
C_2 &=& \lambda_1
{\vec{a}}_{i} \cdot {\vec{b}}_{r} {\vec{b}}^{r} \cdot {\vec{a}}^{i}
+ \lambda_2
{\vec{a}}_{i} \cdot {\vec{\beta}}_{x} {\vec{\beta}}^{x} \cdot {\vec{a}}^{i}
+\lambda_3
{\vec{\alpha}}_{\mu} \cdot {\vec{b}}_{r} {\vec{b}}^{r} \cdot
{\vec{\alpha}}^{\mu}
+ \lambda_4
{\vec{\alpha}}_{\mu} \cdot {\vec{\beta}}_{x} {\vec{\beta}}^{x} \cdot
{\vec{\alpha}}^{\mu} \cr
&+& \lambda_5
{\vec{b}}^{r} \cdot {\vec{a}}^{i} {\vec{a}}_{i} \cdot {\vec{b}}_{r}
+ \lambda_6
{\vec{\beta}}^{x} \cdot {\vec{a}}^{i}{\vec{a}}_{i} \cdot {\vec{\beta}}_{x}
+ \lambda_7
{\vec{b}}^{r} \cdot {\vec{\alpha}}^{\mu}
{\vec{\alpha}}_{\mu} \cdot {\vec{b}}_{r}
+ \lambda_8
{\vec{\beta}}^{x} \cdot {\vec{\alpha}}^{\mu}
{\vec{\alpha}}_{\mu} \cdot {\vec{\beta}}_{x} \cr
&+& \mu_1 L^{i}_{j} L^{j}_{i} + \mu_2 R^{r}_{s} R^{s}_{r} +
\mu_3 A^{\mu}_{\nu} A^{\nu}_{\mu} + \mu_4 B^{x}_{y} B^{y}_{x} \cr
&+& \rho_1 {\vec{a}}^{i} \cdot {\vec{\alpha}}_{\mu}
{\vec{\alpha}}^{\mu} \cdot {\vec{a}}_{i}
+ \rho_2 {\vec{\alpha}}^{\mu} \cdot {\vec{a}}_{i}
{\vec{a}}^{i} \cdot {\vec{\alpha}}_{\mu}
+ \rho_3 {\vec{b}}^{r} \cdot {\vec{\beta}}_{x}
{\vec{\beta}}^{x} \cdot {\vec{b}}_{r}
+ \rho_4
{\vec{\beta}}^{x} \cdot {\vec{b}}_{r} {\vec{b}}^{r} \cdot {\vec{\beta}}_{x} \cr
&+& \sigma_1 E^2 + \sigma_2 C^2 + \sigma_3 (D-F)^2
+ \sigma_4 EC + \sigma_5 E(D-F) + \sigma_6 C(D-F) \cr
&+& {\sigma'}_4 CE
+ {\sigma'}_5 (D-F)E + {\sigma'}_6 (D-F)C
\enn

The requirement that $C_2$ commutes with the generators fixes  the coefficients
up to an overall constant. It turns out that the following commutators 
\eq
[C_2,{\vec{b}}^{t} \cdot {\vec{a}}^{k}]
= [C_2,{\vec{b}}^{t} \cdot {\vec{\alpha}}^{\rho}]
=[C_2,{\vec{\beta}}^{z} \cdot {\vec{a}}^{k}]
= [C_2,{\vec{\beta}}^{z} \cdot {\vec{\alpha}}^{\rho}] =0
\en
determine all the coefficients up to an overall multiplicative 
constant (for which $\sigma_3$ turns out to be a convenient choice)
\eqn
\lambda_1&=&-\lambda_2=-\lambda_3=\lambda_4=\lambda_5=
\lambda_6=\lambda_7=\lambda_8 \cr
&=& -\mu_1=-\mu_2=\mu_3=\mu_4=
\rho_1=-\rho_2=\rho_3=-\rho_4 \cr
 &=& {{4(m+n-p-q)} \over {(m+n)(q+p)} } \sigma_3 \cr
 \sigma_1&=& {{-4mn + (m+n)(p+q)} \over {(m+n)(q+p)} } \sigma_3 \cr
\sigma_2 &=& {{-4pq + (m+n)(p+q)} \over {(m+n)(q+p)} } \sigma_3 \cr
{\sigma'}_4 &=& \sigma_4 =
-{{ (n-m)(q-p)} \over {(m+n)(q+p)} } \sigma_3 \cr
 {\sigma'}_5 &=& \sigma_5 =
-{{ (n-m)} \over {(m+n)} } \sigma_3 \cr
{\sigma'}_6 &=& \sigma_6 =
{{ (q-p)} \over {(q+p)} } \sigma_3
\enn

Choosing $\sigma_3 = {{(m+n)(q+p)} \over 4}$ and
moving all creation operators to the left of annihilation operators
one can write the quadratic Casimir operator of $SU(m,n|p+q)$ as:
\eqn
C_2 &=& [(n-q)^2 - (m-p)^2] [N_a -N_b + N_{\alpha} - N_{\beta}]
+ [N_a -N_b + N_{\alpha} - N_{\beta}]^2 \cr
&+& 2P[(q-n)(N_a + N_{\alpha}) + (p-m)(N_b + N_{\beta})] \cr
&+& P(m-p)(n-q)(m+n-p-q-P) \cr
&+& (m+n-p-q) [2 {\vec{b}}^{r} \cdot {\vec{a}}^{i} {\vec{a}}_{i}
\cdot {\vec{b}}_{r} +
2 {\vec{\beta}}^{x} \cdot {\vec{a}}^{i} {\vec{a}}_{i} \cdot {\vec{\beta}}_{x}
+2 {\vec{b}}^{r} \cdot {\vec{\alpha}}^{\mu} {\vec{\alpha}}_{\mu}
\cdot {\vec{b}}_{r} \cr
&+& 2 {\vec{\beta}}^{x} \cdot {\vec{\alpha}}^{\mu}
{\vec{\alpha}}_{\mu} \cdot {\vec{\beta}}_{x}
-{{a}}^{i}(r') {{a}}^{j}(s') {{a}}_{j}(r') {{a}}_{i}(s')
- {{b}}^{r}(r') {{b}}^{s}(s') {{b}}_{s}(r') {{b}}_{r}(s') \cr
&-& {{\alpha}}^{\mu}(r') {{\alpha}}^{\nu}(s')
{{\alpha}}_{\nu}(r') {{\alpha}}_{\mu}(s')
- {{\beta}}^{x}(r') {{\beta}}^{y}(s')
{{\beta}}_{y}(r') {{\beta}}_{x}(s') \cr
&-& 2{{a}}^{i}(r'){{\alpha}}^{\mu}(s')
{{\alpha}}_{\mu}(r'){{a}}_{i}(s')
-2{{b}}^{r}(r'){{\beta}}^{x}(s'){{\beta}}_{x}(r'){{b}}_{r}(s')]
\enn

Some special cases:

a) If $m+n=p+q$, the quadratic Casimir operator 
 becomes

\eq
C_2 = [N_a -N_b + N_{\alpha} - N_{\beta} + {(m-n+q-p) \over 2} P]^2
\en
where $(N_a -N_b + N_{\alpha} - N_{\beta} + {(m-n+q-p) \over 2} P)$ 
is simply the generator that
commutes with all the generators of $PSU(m,n|p+q)$.
We should note that for $m+n=p+q$ the operator $C_2$ is not a
Casimir operator of $PSU(m,n|p+q)$. The smallest non-trivial
representation of $PSU(m,n|p+q)$ is the adjoint representation
and can not be realized in terms of bilinears of superoscillators 
transforming in the fundamental representation of $SU(m,n|p+q)$.

b) For the case of $P=1$ the most general form (valid for
arbitrary $m, n, p, q$) of $C_2$ reduces to

\eqn
C_2 &=& (m+n-p-q-1)[(n-q-m+p)(N_a -N_b + N_{\alpha} - N_{\beta}) \cr
&-& (N_a -N_b + N_{\alpha} - N_{\beta})^2 +(m-p)(n-q)]
\enn

\section{Unitary Supermultiplets of $SU(2,2|4)$}
\setcounter{equation}{0}

The UIRs of $SU(2,2|4)$ are constructed following the
general procedure outlined in the introduction. To this end we 
let the indices $i,j,..;r,s,..;\mu, \nu,..; x,y,..$ run from
1 to 2. 
Starting from the ground state $ |\Omega \rangle$
in the Fock space transforming irreducibly under $SU(2|2) \times
SU(2|2) \times U(1)$ and is annihilated by $L^{-}$ (given
in terms of oscillators $\xi$ and $\eta$ as in eq. (7.7)), one can
generate the UIRs of $SU(2,2|4)$ by repeated application of
$L^{+}$ 
\eq
 |\Omega \rangle ,\quad  L^{+1}|\Omega \rangle ,\quad
L^{+1} L^{+1}|\Omega \rangle , ...
\en
The irreducibility of UIRs of $SU(2,2|4)$ follows from the
irreducibility of $|\Omega \rangle$ under $SU(2|2) \times
SU(2|2) \times U(1)$.

When restricted to the subspace involving purely bosonic
oscillators we get the subalgebra $SU(2,2)$ and the above 
construction yields its positive energy UIR's. 
Similarly, when restricted to the subspace involving purely
fermionic oscillators we get the compact internal 
symmetry group $ SU(4)$
Then the above construction yields the representations of $SU(4)$ in
its $SU(2)\times SU(2) \times U(1)$ basis. 

The positive energy UIR's of $SU(2,2|4)$ decompose into a direct sum of
finitely many positive energy UIR's of $SU(2,2)$ 
transforming in certain representations
of the internal symmetry group $SU(4)$. 
Thus each positive energy UIR of $SU(2,2|4)$
corresponds to a supermultiplet of fields living in $AdS_5$ or its boundary.
The bosonic and fermionic fields in $AdS_5$ or its boundary will be denoted
as $\Phi_{(j_{L},j_{R})}(E)$ and $\Psi_{(j_{L},j_{R})}(E)$, respectively.

We shall assume that the vacuum vector $|0\rangle$ is CPT invariant and
we shall call the supermultiplets defined by taking $|0\rangle $ as the lowest 
weight vector the CPT self-conjugate supermultiplets. Furthermore, we shall
refer to the supermultiplets obtained by the interchange of $\xi$ and $\eta$
type superoscillators as conjugate supermultiplets.

\subsection{Doubleton Supermultiplets of $SU(2,2|4)$}
\setcounter{equation}{0}

More explicitly, consider $|0\rangle$ as
the lowest weight vector of $SU(2,2|4)$. $|0\rangle$ is
automatically a lowest weight vector of $SU(2,2) \times U(4)$.
By acting on $|0\rangle$ with the supersymmetry 
generators ${\vec{a}}^{i} \cdot {\vec{\beta}}^{x}$ and
${\vec{b}}^{r} \cdot {\vec{\alpha}}^{\mu}$, one generates 
additional lowest weight vectors of $SU(2,2) \times U(4)$. The 
action of $L^{ir}$ on these lowest weight vectors generates the 
higher Fourier modes of the corresponding fields
 and the action of $L^{\mu x}$ corresponds to moving within the
 respective $SU(4)$ representation.
 We find that the AdS scalar fields corresponding to
$|0\rangle$ transform as ${\bf 6}$ of $SU(4)$, 
the spinor fields corresponding to $L^{ix} |0\rangle$ and $L^{r \mu} |0\rangle$
transform respectively as ${\bf \bar{4}}$ and ${\bf 4}$ of $SU(4)$
 and finally, the self-dual and anti-self-dual tensor fields corresponding to
$L^{ix} L^{jy} |0\rangle$ and $L^{r \mu} L^{s \nu} |0\rangle$ transform
as two singlets of $SU(4)$.

This supermultiplet is the  CPT self-conjugate 
doubleton supermultiplet (we have
used a pair of oscillators in its construction) and it is the supermultiplet
of ${\cal{N}}=4$ supersymmetric Yang-Mills theory in 4-d \cite{mgnm}.
The contents of this particular doubleton supermultiplet are given in
Table 1. (We will continue to use this form of representing our
results in what follows.)

\vspace{.2cm}
\begin{center}
\begin{tabular}{|c|c|c|c|c|c|}
\hline
$SU(2,2) \times SU(4)$ lwv &E &($j_{L},j_{R}$) & SU(4) &$Y$ &Field
\\ \hline
$|0\rangle $          &1 & (0,0)     & 6   & 0 &$\Phi_{0,0}$
\\ \hline
$a^i \beta^x |0\rangle$   &3/2  & (1/2,0) & $\bar{4}$ & -1 &$\Psi_{1/2,0}$
\\ \hline
$b^r \alpha^{\mu}  |0\rangle$        &3/2& (0,1/2)   & 4  &1 &$\Psi_{0,1/2}$
\\ \hline
$a^i a^j  \beta^x \beta^y |0\rangle$ &2& (1,0)  & $\bar{1}$& -2 &$\Phi_{1,0}$
\\ \hline
$b^r b^s \alpha^{\mu} \alpha^{\nu} |0\rangle$   &2& (0,1) & 1 & 2 &$\Phi_{0,1}$
\\ \hline
\end{tabular}
\end{center}
Table 1. The doubleton supermultiplet
corresponding to the lwv $|\Omega \rangle = |0\rangle$.
The first column indicates the lowest weight vectors (lwv) of
$SU(2,2) \times SU(4)$.
Also, $Y= N_{\alpha} - N_{\beta}; E= (N_a +N_b)/2 +P \equiv N/2 +1$.
$\Phi $ and $\Psi$ denote bosonic and fermionic fields respectively. 
\vspace{.7cm}

Note, that by insisting on
CPT self-conjugacy of the irreducible supermultiplet
(which means, in this case, that
$\Omega = |0\rangle$) we get only the above
${\cal{N}} =4$ Yang-Mills supermultiplet.
But there are other irreducible doubleton supermultiplets which
are not CPT self-conjugate, and which are different from the 4-d
${\cal{N}}=4$ supersymmetric Yang-Mills supermultiplet.

If we take
\eq
|\Omega \rangle = \xi^A |0\rangle \equiv a^i |0\rangle
\oplus \alpha^{\mu} |0\rangle
= |{\sonebox}, 1\rangle
\en
we get the supermultiplet represented in Table 2. (See the appendix for a
quick review of the supertableaux notation \cite{bbars}.)

\vspace{.2cm}
\begin{center}
\begin{tabular}{|c|c|c|c|c|c|}
\hline
$SU(2,2) \times SU(4)$ lwv   &E  & ($j_{L},j_{R}$)& SU(4)
&$Y$ &Field
\\ \hline
$a^i|0\rangle $                        &3/2& (1/2,0)  & 6   & 0 &$\Psi_{1/2,0}$
\\ \hline
$a^i a^j \beta^x |0\rangle$ &2  & (1,0)  & $\bar{4}$& -1 &$\Phi_{1,0}$
\\ \hline
$a^i a^j a^k \beta^x \beta^y |0\rangle$&5/2&(3/2,0)&
$\bar{1}$&-2&$\Psi_{3/2,0}$
\\ \hline
$\alpha^{\mu} |0\rangle$               &1  & (0,0) & 4 &1 &$\Phi_{0,0}$
\\ \hline
$b^r \alpha^{\mu} \alpha^{\nu} |0\rangle$  &3/2& (0,1/2)  & 1& 2&$\Psi_{0,1/2}$
\\ \hline
\end{tabular}
\end{center}
Table 2. The doubleton supermultiplet
corresponding to the lwv $|\Omega\rangle=\xi^A |0\rangle
= |{\sonebox},1 \rangle  $. 
\vspace{.7cm}

The  conjugate supermultiplet to the one above can be obtained if
we take
\eq
|\Omega \rangle = \eta^A |0\rangle \equiv b^r |0\rangle
\oplus \beta^{x} |0\rangle
= |1,{\sonebox} \rangle
\en
Then we get the supermultiplet represented in Table 3.

\vspace{.2cm}
\begin{center}
\begin{tabular}{|c|c|c|c|c|c|}
\hline
$SU(2,2) \times SU(4)$ lwv &E  & ($j_{L},j_{R}$)& SU(4) &$Y$ &Field
\\ \hline
$b^r|0\rangle $  &3/2& (0,1/2)        & 6  & 0 &$\Psi_{0,1/2}$
\\ \hline
$b^r b^s \alpha^{\mu} |0\rangle$ &2  & (0,1)& ${4}$& 1 &$\Phi_{0,1}$
\\ \hline
$b^r b^s b^t \alpha^{\mu} \alpha^{\nu}|0\rangle $ &5/2& (0,3/2)  & 1   & 2
&$\Psi_{0,3/2}$
\\ \hline
$\beta^x |0\rangle$      &1  & (0,0) & $\bar{4}$  &-1 &$\Phi_{0,0}$
\\ \hline
$a^i \beta^x \beta^y |0\rangle$  &3/2& (1/2,0) & $\bar{1}$   & -2
&$\Psi_{1/2,0}$
\\ \hline
\end{tabular}
\end{center}
Table 3. The doubleton supermultiplet
corresponding to the lwv $|\Omega\rangle=\eta^A |0\rangle
= |1,{\sonebox} \rangle  $ . 
\vspace{.7cm}

The above supermultiplets (Table 2. and 3.) have spin range $3/2$.

Similarly by taking
\eq
|\Omega \rangle = \xi^A \xi^B |0\rangle \equiv a^i a^j |0\rangle
\oplus a^i \alpha^{\mu} |0\rangle
 \oplus \alpha^{\mu} \alpha^{\nu} |0\rangle
= |{\stwobox}, 1 \rangle
\en
we get the supermultiplet represented in Table 4.

\vspace{.2cm}
\begin{center}
\begin{tabular}{|c|c|c|c|c|c|}
\hline
$SU(2,2) \times SU(4) $ lwv    &E  & ($j_{L},j_{R}$)& SU(4)
&$Y$ &Field
\\ \hline
$a^i a^j|0\rangle $      &2& (1,0)  & 6   & 0 &$\Phi_{1,0}$
\\ \hline
$a^i a^j a^k \beta^x |0\rangle$ &5/2  & (3/2,0)  & $\bar{4}$& -1 &$\Psi_{3/2,0}$
\\ \hline
$a^i a^j a^k a^l \beta^x \beta^y |0\rangle$
&3&(2,0)& $\bar{1}$&-2&$\Phi_{2,0}$
\\ \hline
$a^i \alpha^{\mu} |0\rangle$       &3/2  & (1/2,0) & 4 &1 &$\Psi_{1/2,0}$
\\ \hline
$\alpha^{\mu} \alpha^{\nu} |0\rangle$  &1& (0,0)  & 1& 2&$\Phi_{0,0}$
\\ \hline
\end{tabular}
\end{center}
Table 4. The doubleton supermultiplet
corresponding to the lwv $|\Omega\rangle=\xi^A \xi^B |0\rangle
= |{\stwobox}, 1 \rangle  $ . 
\vspace{.7cm}

The conjugate irreducible supermultiplet is obtained by taking
\eq
|\Omega \rangle = \eta^A \eta^B |0\rangle \equiv b^r b^s |0\rangle
\oplus b^r \beta^{x} |0\rangle
 \oplus \beta^{x} \beta^{y} |0\rangle
= |1, {\stwobox} \rangle
\en
The result is given in Table 5.

\vspace{.2cm}
\begin{center}
\begin{tabular}{|c|c|c|c|c|c|}
\hline
$SU(2,2) \times SU(4) $ lwv &E  & ($j_{L},j_{R}$)& SU(4) &$Y$
 &Field
\\ \hline
$b^r b^s|0\rangle $  &2& (0,1)        & 6  & 0 &$\Phi_{0,1}$
\\ \hline
$b^r b^s b^t\alpha^{\mu} |0\rangle$ &5/2  & (0,3/2)& ${4}$& 1 &$\Psi_{0,3/2}$
\\ \hline
$b^r b^s b^t b^u \alpha^{\mu} \alpha^{\nu}|0\rangle $ &3& (0,2)  & 1   & 2
&$\Phi_{0,2}$
\\ \hline
$b^r \beta^x |0\rangle$      &3/2  & (0,1/2) & $\bar{4}$  &-1 &$\Psi_{0,1/2}$
\\ \hline
$\beta^x \beta^y |0\rangle$  &1& (0,0) & $\bar{1}$   & -2
&$\Phi_{0,0}$
\\ \hline
\end{tabular}
\end{center}
Table 5. The doubleton supermultiplet
corresponding to the lwv $|\Omega\rangle=\eta^A \eta^B |0\rangle
= |1, {\stwobox} \rangle  $ . 
\vspace{.7cm}

The direct sum of the supermultiplets defined by the lowest
weight vectors $\xi^A \xi^B |0\rangle$ and
$\eta^A \eta^B |0\rangle$ is parity invariant and corresponds to the 
${\cal{N}}=4$
conformal supergravity multiplet in $d=4$. 

In general, we could take 
\eqn
|\Omega \rangle &=& \xi^{A_1} \xi^{A_2}...\xi^{A_{2j}} |0\rangle \cr
|\Omega \rangle &=& \eta^{A_1} \eta^{A_2}...\eta^{A_{2j}} |0\rangle
\enn
as our lowest weight vectors.

For $j \geq 1$, the general doubleton supermultiplets,
obtained by taking
\eq
|\Omega \rangle = \xi^{A_1} \xi^{A_2}...\xi^{A_{2j}} |0\rangle
= |\underbrace{\sgenrowbox}_{2j}, 1 \rangle
\en
are represented in Table 6.

\vspace{.2cm}
\begin{center}
\begin{tabular}{|c|c|c|c|c|}
\hline
E                 & ($j_{L},j_{R}$)     & SU(4)     & $U(1)_{Y}$
& Field\\ \hline
j+1               & (j,0)               & 6         & 0 &$\Phi_{j,0}$
\\ \hline
j+{3/2}           & (j+1/2,0)           & $\bar{4}$ & -1 &$\Psi_{j+1/2,0}$
\\ \hline
j+{1/2}           & (j-1/2,0)           & 4         & 1 &$\Psi_{j-1/2,0}$
\\ \hline
j+ 2              & (j+1 ,0)            & $\bar{1}$ & -2 &$\Phi_{j+1,0}$
\\ \hline
j                 & (j-1,0)             & 1         & 2 &$\Phi_{j-1,0}$
\\ \hline
\end{tabular}
\end{center}
Table 6. The doubleton supermultiplet
corresponding to the lwv 
$|\Omega \rangle = \xi^{A_1} \xi^{A_2}...\xi^{A_{2j}} |0\rangle
= |\underbrace{\sgenrowbox}_{2j}, 1 \rangle $.
\vspace{.7cm}

Here we assume that $j$ takes integer values. For $j$ half-integer the roles
of $\Phi$ and $\Psi$ are reversed.

The conjugate supermultiplets to the ones above are obtained by taking
\eq
|\Omega \rangle = \eta^{A_1} \eta^{A_2}...\eta^{A_{2j}} |0\rangle
= |1, \underbrace{\sgenrowbox}_{2j} \rangle
\en
and have the form represented in Table 7.

\vspace{.2cm}
\begin{center}
\begin{tabular}{|c|c|c|c|c|}
\hline
E                 & ($j_{L},j_{R}$)     & SU(4)     & $U(1)_{Y}$
&Field\\ \hline
j+1               & (0,j)               & 6         & 0 &$\Phi_{0,j}$
\\ \hline
j+{3/2}           & (0,j+1/2)           & 4         & 1 &$\Psi_{0,j+1/2}$
\\ \hline
j+{1/2}           & (0,j-1/2)           & $\bar{4}$ & -1 &$\Psi_{0,j-1/2}$
\\ \hline
j+ 2              & (0,j+1)             & 1         & 2 &$\Phi_{0,j+1}$
\\ \hline
j                 & (0,j-1)             & $\bar{1}$ & -2 &$\Phi_{0,j-1}$
\\ \hline
\end{tabular}
\end{center}
Table 7. The doubleton supermultiplet
corresponding to the lwv 
$|\Omega \rangle = \eta^{A_1} \eta^{A_2}...\eta^{A_{2j}} |0\rangle
= |1, \underbrace{\sgenrowbox}_{2j} \rangle $.
\vspace{.7cm}

\section{"Massless" Irreducible Representations of $SU(2,2|4)$}
\setcounter{equation}{0}

The
"massless" supermultiplets 
of $SU(2,2|4)$ are obtained by tensoring two doubleton
supermultiplets, i.e. by taking $P=2$.
In this case
the vacuum $|0\rangle$ taken as the lowest weight vector of $SU(2,2|4)$
leads to the "massless" 
${\cal{N}}=8$ graviton supermultiplet in $AdS_5$ \cite{mgnm}.
It is a short supermultiplet having spin range 2.
However, generic "massless" supermultiplets
have spin range 4. We also have "massless" supermultiplets that are
short in addition to the graviton supermultiplet.

For example,
consider the "massless" supermultiplet ($P \equiv 2$) defined by
the lowest weight vector
\eq
| \Omega \rangle =
\xi^{A}(2) |0 \rangle \equiv a^i(2) |0\rangle \oplus \alpha^{\mu}(2) |0\rangle
= |{\sonebox}, 1 \rangle
\en

The application of the susy generators ${\vec{a}}^i \cdot {\vec{\beta}}^x$ and
${\vec{b}}^r \cdot {\vec{\alpha}}^{\mu}$
leads to the "massless" supermultiplet represented in Table 8. The possible
$SU(4)$ representations and the corresponding lowest weight vectors for
$P=2$ are given in the appendix.

\vspace{.2cm}
\begin{center}
\begin{tabular}{|c|c|c|c|c|}
\hline
E  & ($j_{L},j_{R}$)     & SU(4) Dynkin   & $U(1)_{Y}$
&Field \\ \hline
2       &(0,0)  &(1,1,0)  &1  &$\Phi_{0,0}$\\  \hline
5/2               & (1/2,0)   & (0,2,0) & 0 &$\Psi_{1/2,0}$\\ \hline
5/2   &(1/2,0)  &(1,0,1)  &0  &$\Psi_{1/2,0}$\\ \hline
5/2   &(0,1/2)  &(0,1,0)  &2  &$\Psi_{0,1/2}$\\  \hline
5/2  &(0,1/2)  &(2,0,0)  &2  &$\Psi_{0,1/2}$\\ \hline
3   &(1,0)  &(1,0,0)  &-1  &$\Phi_{1,0}$\\  \hline
3  &(0,1)  &(1,0,0)  &3  &$\Phi_{0,1}$\\ \hline
3  &(0,0)  &(1,0,0)  &3  &$\Phi_{0,0}$\\   \hline
3           & (0,0)           & (0,1,1)         & -1 &$\Phi_{0,0}$\\ \hline
3           & (1,0)           & (0,1,1) & -1 &$\Phi_{1,0}$\\ \hline
3              & (1/2,1/2)      & (1,1,0)    & 1 &$\Phi_{1/2,1/2}$\\ \hline
3    &(1/2,1/2)  &(0,0,1) & 1 &$\Phi_{1/2,1/2}$\\ \hline
7/2      &(3/2,0) &(0,1,0)  &-2 &$\Psi_{3/2,0}$\\ \hline
7/2   &(1/2,0)  &(0,0,2)  &-2  &$\Psi_{1/2,0}$\\ \hline
7/2  &(1/2,0)  &(0,1,0)  &-2  &$\Psi_{1/2,0}$\\ \hline
7/2  &(1/2,1)  &(0,1,0)  &2  &$\Psi_{1/2,1}$\\ \hline
7/2  &(1,1/2)  &(1,0,1)  &0  &$\Psi_{1,1/2}$\\ \hline
7/2  &(0,1/2)  &(0,0,0)  &4  &$\Psi_{0,1/2}$\\  \hline
7/2 &(1,1/2) &(0,0,0) &0  &$\Psi_{1,1/2}$\\ \hline
4 &(1,0)  &(0,0,1)  &-3  &$\Phi_{1,0}$\\ \hline
4 &(0,0) &(0,0,1)  & -3 &$\Phi_{0,0}$\\ \hline
4 &(3/2,1/2) & (1,0,0) &-1 &$\Phi_{3/2,1/2}$\\ \hline
4  &(1,1)  &(0,0,1)  &1  &$\Phi_{1,1}$\\  \hline
9/2 &(1/2,0)  &(0,0,0)  &-4  &$\Psi_{1/2,0}$\\ \hline
9/2  &(3/2,1)  &(0,0,0)  &0 &$\Psi_{3/2,1}$\\  \hline
\end{tabular}
\end{center}
Table 8. The "massless" supermultiplet
corresponding to the lwv 
$| \Omega \rangle = \xi^{A}(2) |0 \rangle \equiv a^i(2) |0\rangle 
\oplus \alpha^{\mu}(2) |0\rangle
= |{\sonebox}, 1 \rangle  $. 
\vspace{.7cm}

Similarly, consider the "massless" supermultiplet defined by
\eq
|\Omega \rangle =
\eta^{A}(2) |0 \rangle \equiv b^r(2) |0\rangle \oplus \beta^{x}(2) |0\rangle
= |1, {\sonebox} \rangle
\en

Again, the application of the susy generators
${\vec{a}}^i \cdot {\vec{\beta}}^x$ and
${\vec{b}}^r \cdot {\vec{\alpha}}^{\mu}$
leads to the "massless" supermultiplet represented in Table 9.

\vspace{.2cm}
\begin{center}
\begin{tabular}{|c|c|c|c|c|}
\hline
E                 & ($j_{L},j_{R}$)     & SU(4) Dynkin   & $U(1)_{Y}$
&Field\\ \hline
2       &(0,0)  &(0,1,1)  &-1  &$\Phi_{0,0}$\\  \hline
5/2               & (0,1/2)           & (0,2,0)   & 0 &$\Psi_{0,1/2}$\\ \hline
5/2   &(0,1/2)  &(1,0,1)  &0  &$\Psi_{0,1/2}$\\ \hline
5/2   &(1/2,0)  &(0,1,0)  &-2  &$\Psi_{1/2,0}$\\  \hline
5/2  &(1/2,0)  &(0,0,2)  &-2  &$\Psi_{1/2,0}$\\ \hline
3   &(0,1)  &(0,0,1)  &1  &$\Phi_{0,1}$\\  \hline
3  &(1,0)  &(0,0,1)  &-3  &$\Phi_{1,0}$\\ \hline
3  &(0,0)  &(0,0,1)  &-3  &$\Phi_{0,0}$\\   \hline
3           & (0,0)           & (1,1,0)         & 1 &$\Phi_{0,0}$\\ \hline
3           & (0,1)           & (1,1,0) & 1 &$\Phi_{0,1}$\\ \hline
3              & (1/2,1/2)      & (0,1,1)    & -1 &$\Phi_{1/2,1/2}$\\ \hline
3    &(1/2,1/2)  &(1,0,0) & -1 &$\Phi_{1/2,1/2}$\\ \hline
7/2      &(0,3/2) &(0,1,0)  &2 &$\Psi_{0,3/2}$\\ \hline
7/2   &(0,1/2)  &(2,0,0)  &2  &$\Psi_{0,1/2}$\\ \hline
7/2  &(0,1/2)  &(0,1,0)  &2  &$\Psi_{0,1/2}$\\ \hline
7/2  &(1,1/2)  &(0,1,0)  &-2  &$\Psi_{1,1/2}$\\ \hline
7/2  &(1/2,1)  &(1,0,1)  &0  &$\Psi_{1/2,1}$\\ \hline
7/2  &(1/2,0)  &(0,0,0)  &-4  &$\Psi_{1/2,0}$\\  \hline
7/2 &(1/2,1) &(0,0,0) &0  &$\Psi_{1/2,1}$\\ \hline
4 &(0,1)  &(1,0,0)  &3  &$\Phi_{0,1}$\\ \hline
4 &(0,0) &(1,0,0)  & 3 &$\Phi_{0,0}$\\ \hline
4 &(1/2,3/2) & (0,0,1) &1 &$\Phi_{1/2,3/2}$\\ \hline
4  &(1,1)  &(1,0,0)  &-1  &$\Phi_{1,1}$\\  \hline
9/2 &(0,1/2)  &(0,0,0)  &4  &$\Psi_{0,1/2}$\\ \hline
9/2  &(1,3/2)  &(0,0,0)  &0 &$\Psi_{1,3/2}$\\  \hline
\end{tabular}
\end{center}
Table 9. The "massless" supermultiplet
corresponding to the lwv 
$|\Omega \rangle = \eta^{A}(2) |0 \rangle \equiv b^r(2) |0\rangle 
\oplus \beta^{x}(2) |0\rangle
= |1, {\sonebox} \rangle  $ .
\vspace{.7cm}

The above supermultiplets have spin range $5/2$.

The general form of the "massless" supermultiplet that is obtained from
\eq
|\Omega \rangle = \xi^{A_1}(1) \xi^{A_2}(1) ... \xi^{A_{2j}}(1) |0\rangle
= |\underbrace{\sgenrowbox}_{2j}, 1 \rangle
\en
is represented in Table 10 (we take $j>3/2$).

\vspace{.2cm}
\begin{center}
\begin{tabular}{|c|c|c|c|c|}
\hline
E                 & ($j_{L},j_{R}$)     & SU(4) Dynkin   & $U(1)_{Y}$
&Field\\ \hline
j+1  &(j-1,0) &(0,1,0)  &2  &$\Phi_{j-1,0}$\\ \hline
j+3/2  &(j-3/2,0)  &(0,0,1)  &1   &$\Psi_{j-3/2,0}$\\  \hline
j+3/2  &(j-1/2,0)  &(0,0,1)  &1  &$\Psi_{j-1/2,0}$\\  \hline
j+3/2  &(j-1,1/2)  &(1,0,0)  &3  &$\Psi_{j-1,1/2}$\\  \hline
j+3/2 &(j-1/2,0)  &(1,1,0)  &1  &$\Psi_{j-1/2,0}$\\  \hline
j+2  &(j-1,0)  &(0,0,0)  &0   &$\Phi_{j-1,0}$\\ \hline
j+2  &(j-1/2,1/2)  &(0,1,0)  &2   &$\Phi_{j-1/2,1/2}$\\ \hline
j+2  &(j,0)  &(1,0,1)  &0   &$\Phi_{j,0}$\\ \hline
j+2  &(j-1,0)  &(1,0,1)  &0  &$\Phi_{j-1,0}$\\ \hline
j+2  &(j-1/2,1/2)  &(2,0,0)  &2  &$\Phi_{j-1/2,1/2}$\\ \hline
j+2 &(j,0) & (0,2,0)  & 0 &$\Phi_{j,0}$\\ \hline
j+2  &(j,0)  &(0,0,0)  &0  &$\Phi_{j,0}$\\  \hline
j+2  &(j-2,0)  &(0,0,0)  &0   &$\Phi_{j-2,0}$\\  \hline
j+2  &(j-1,1)  &(0,0,0)  &4  &$\Phi_{j-1,1}$\\  \hline
j+5/2 & (j+1/2,0)  & (0,1,1)  & -1 &$\Psi_{j+1/2,0}$\\ \hline
j+5/2  &(j-1/2,0)  &(0,1,1)  &-1  &$\Psi_{j-1/2,0}$\\  \hline
j+5/2  &(j,1/2)  & (1,1,0)  &1  &$\Psi_{j,1/2}$\\ \hline
j+5/2  &(j-1/2,1)  &(1,0,0)  &3  &$\Psi_{j-1/2,1}$\\  \hline
j+5/2  &(j+1/2,0)  &(1,0,0)  &-1  &$\Psi_{j+1/2,0}$\\  \hline
j+5/2  &(j-3/2,0)  &(1,0,0)  & -1  &$\Psi_{j-3/2,0}$\\  \hline
j+5/2  &(j-1/2,0)  &(1,0,0)  & -1  &$\Psi_{j-1/2,0}$\\  \hline
j+3  &(j+1,0) & (0,1,0)  &-2  &$\Phi_{j+1,0}$\\ \hline
j+3  &(j,0) & (0,0,2)  &-2  &$\Phi_{j,0}$\\ \hline
j+3  &(j,0) & (0,1,0)  &-2  &$\Phi_{j,0}$\\ \hline
j+3  &(j-1,0) &(0,1,0)  &-2  &$\Phi_{j-1,0}$\\ \hline
j+3  &(j,1) & (0,1,0)  &2  &$\Phi_{j,1}$\\ \hline
j+3  &(j+1/2,1/2)  &(1,0,1)  &0  &$\Phi_{j+1/2,1/2}$\\ \hline
j+7/2  &(j+1,1/2) &(1,0,0)  &-1  &$\Psi_{j+1,1/2}$\\  \hline
j+7/2  &(j+1/2,1)  &(0,0,1)  &1  &$\Psi_{j+1/2,1}$\\  \hline
j+7/2  &(j+1/2,0)  &(0,0,1)  & -3  &$\Psi_{j+1/2,0}$ \\ \hline
j+7/2  &(j-1/2,0)  &(0,0,1)  &-3  &$\Psi_{j-1/2,0}$\\ \hline
j+4  &(j,0)  &(0,0,0)  &-4  &$\Phi_{j,0}$\\  \hline
j+4  &(j+1,1) &(0,0,0)  & 0  &$\Phi_{j+1,1}$\\  \hline
\end{tabular}
\end{center}
Table 10. The "massless" supermultiplet
corresponding to the lwv   
$|\Omega \rangle = \xi^{A_1}(1) \xi^{A_2}(1) ... \xi^{A_{2j}}(1) |0\rangle
= |\underbrace{\sgenrowbox}_{2j}, 1 \rangle $.
\vspace{.7cm}

Note that $j$ is again assumed to take only integer values in Table 10.
For $j$ half-integer, $\Phi$ and $\Psi$ should be interchanged.

On the other hand,
the general form of the "massless" supermultiplet that is obtained from
\eq
|\Omega \rangle = \eta^{A_1}(1) \eta^{A_2}(1) ... \eta^{A_{2j}}(1) |0\rangle
= |1, \underbrace{\sgenrowbox}_{2j} \rangle
\en
is represented in Table 11. ($j>3/2$)

\vspace{.2cm}
\begin{center}
\begin{tabular}{|c|c|c|c|c|}
\hline
E                 & ($j_{L},j_{R}$)     & SU(4) Dynkin   & $U(1)_{Y}$
&Field \\ \hline
j+1  &(0,j-1) &(0,1,0)  &-2  &$\Phi_{0,j-1}$\\ \hline
j+3/2  &(0,j-3/2)  &(1,0,0)  &-1   &$\Psi_{0,j-3/2}$\\  \hline
j+3/2  &(0,j-1/2)  &(1,0,0)  &-1  &$\Psi_{0,j-1/2}$\\  \hline
j+3/2  &(1/2,j-1)  &(0,0,1)  &-3  &$\Psi_{1/2,j-1}$\\  \hline
j+3/2 &(0,j-1/2)  &(0,1,1)  &-1  &$\Psi_{0,j-1/2}$\\  \hline
j+2  &(0,j-1)  &(0,0,0)  &0   &$\Phi_{0,j-1}$\\ \hline
j+2  &(1/2,j-1/2)  &(0,1,0)  &-2   &$\Phi_{1/2,j-1/2}$\\ \hline
j+2  &(0,j)  &(1,0,1)  &0   &$\Phi_{0,j}$\\ \hline
j+2  &(0,j-1)  &(1,0,1)  &0  &$\Phi_{0,j-1}$\\ \hline
j+2  &(1/2,j-1/2)  &(0,0,2)  &-2  &$\Phi_{1/2,j-1/2}$\\ \hline
j+2 &(0,j) & (0,2,0)  & 0 &$\Phi_{0,j}$\\ \hline
j+2  &(0,j)  &(0,0,0)  &0  &$\Phi_{0,j}$\\  \hline
j+2  &(0,j-2)  &(0,0,0)  &0   &$\Phi_{0,j-2}$\\  \hline
j+2  &(1,j-1)  &(0,0,0)  &-4  &$\Phi_{1,j-1}$\\  \hline
j+5/2 & (0,j+1/2)  & (1,1,0)  & 1 &$\Psi_{0,j+1/2}$\\ \hline
j+5/2  &(0,j-1/2)  &(1,1,0)  &1  &$\Psi_{0,j-1/2}$\\  \hline
j+5/2  &(1/2,j)  & (0,1,1)  &-1  &$\Psi_{1/2,j}$\\ \hline
j+5/2  &(1,j-1/2)  &(0,0,1)  &-3  &$\Psi_{1,j-1/2}$\\  \hline
j+5/2  &(0,j+1/2)  &(0,0,1)  &1  &$\Psi_{0,j+1/2}$\\  \hline
j+5/2  &(0,j-3/2)  &(0,0,1)  & 1  &$\Psi_{0,j-3/2}$\\  \hline
j+5/2  &(0,j-1/2)  &(0,0,1)  & 1  &$\Psi_{0,j-1/2}$\\  \hline
j+3  &(0,j+1) & (0,1,0)  &2  &$\Phi_{0,j+1}$\\ \hline
j+3  &(0,j) & (2,0,0)  &2  &$\Phi_{0,j}$\\ \hline
j+3  &(0,j) & (0,1,0)  &2  &$\Phi_{0,j}$\\ \hline
j+3  &(0,j-1) &(0,1,0)  &2  &$\Phi_{0,j-1}$\\ \hline
j+3  &(1,j) & (0,1,0)  &-2  &$\Phi_{1,j}$\\ \hline
j+3  &(1/2,j+1/2)  &(1,0,1)  &0  &$\Phi_{1/2,j+1/2}$\\ \hline
j+7/2  &(1/2,j+1) &(0,0,1)  &1  &$\Psi_{1/2,j+1}$\\  \hline
j+7/2  &(1,j+1/2)  &(1,0,0)  &-1  &$\Psi_{1,j+1/2}$\\  \hline
j+7/2  &(0,j+1/2)  &(1,0,0)  & 3   &$\Psi_{0,j+1/2}$\\ \hline
j+7/2  &(0,j-1/2)  &(1,0,0)  &3  &$\Psi_{0,j-1/2}$\\ \hline
j+4  &(0,j)  &(0,0,0)  &4  &$\Phi_{0,j}$\\  \hline
j+4  &(1,j+1) &(0,0,0)  & 0  &$\Phi_{1,j+1}$\\  \hline
\end{tabular}
\end{center}
Table 11. The "massless" supermultiplet
corresponding to the lwv     
$|\Omega \rangle = \eta^{A_1}(1) \eta^{A_2}(1) ... \eta^{A_{2j}}(1) |0\rangle
= |1, \underbrace{\sgenrowbox}_{2j} \rangle $.
\vspace{.7cm}

As before, $j$ is assumed to take only integer values in Table 11.

Finally we list another allowed irreducible
 "massless" supermultiplet which can be obtained from
the following lowest weight vector
\eq
|\Omega \rangle = \xi^{A_1}(1) \xi^{A_2}(1) ... \xi^{A_{2j_{L}}}(1)
\eta^{B_1}(2) \eta^{B_2}(2) ... \eta^{B_{2j_{R}}} (2)|0\rangle =
|\underbrace{\sgenrowbox}_{2j_{L}}, \underbrace{\sgenrowbox}_{2j_{R}} \rangle
\en

\vspace{.2cm}
\begin{center}
\begin{tabular}{|c|c|c|c|c|}
\hline
E                 & ($j_{L},j_{R}$)     & SU(4) Dynkin   & $U(1)_{Y}$
&Field\\ \hline
$j_{L}+j_{R}$ &$(j_{L}-1,j_{R}-1)$ &(0,0,0)  &0  &$\Phi_{j_{L}-1,j_{R}-1}$ 
\\ \hline
$j_{L}+j_{R}+1/2$ &$(j_{L}-1/2,j_{R}-1)$ & (1,0,0) &-1 
&$\Psi_{j_{L}-1/2,j_{R}-1}$
\\ \hline
$j_{L}+j_{R}+1/2$ &$(j_{L}-1,j_{R}-1/2)$ & (0,0,1) &1 
&$\Psi_{j_{L}-1,j_{R}-1/2}$
\\ \hline
$j_{L} + j_{R}+1$  &$(j_{L}-1/2, j_{R}-1/2)$ &(0,0,0)  &0  
&$\Phi_{j_{L}-1/2,j_{R}-1/2}$
\\ \hline
$j_{L} + j_{R}+1$  &$(j_{L}-1, j_{R})$ &(0,1,0)  &2  &$\Phi_{j_{L}-1,j_{R}}$
\\ \hline
$j_{L} + j_{R}+1$  &$(j_{L}, j_{R}-1)$ &(0,1,0)  &-2  &$\Phi_{j_{L},j_{R}-1}$
\\ \hline
$j_{L} + j_{R}+1$  &$(j_{L}-1/2, j_{R}-1/2)$ &(1,0,1)  &0  
&$\Phi_{j_{L}-1/2,j_{R}-1/2}$
\\ \hline
$j_{L} + j_{R}+3/2$  &$(j_{L},j_{R}-1/2)$  &(1,0,0)  &-1   
&$\Psi_{j_{L},j_{R}-1/2}$
\\ \hline
$j_{L} + j_{R}+3/2$  &$(j_{L}-1/2,j_{R})$  &(0,0,1)  &1   
&$\Psi_{j_{L}-1/2,j_{R}}$
\\ \hline
$j_{L} + j_{R}+3/2$  &$(j_{L}+1/2,j_{R}-1)$  &(0,0,1)  &-3   
&$\Psi_{j_{L}+1/2,j_{R}-1}$
\\  \hline
$j_{L}+j_{R}+3/2$  &$(j_{L}-1,j_{R}+1/2)$  &(1,0,0)  &3  
&$\Psi_{j_{L}-1,j_{R}+1/2}$
\\  \hline
$j_{L} + j_{R}+3/2$  &$(j_{L},j_{R}-1/2)$  &(0,1,1)  &-1  
&$\Psi_{j_{L},j_{R}-1/2}$
\\  \hline
$j_{L}+j_{R}+3/2$ &$(j_{L}-1/2,j_{R})$  &(1,1,0)  &1  &$\Psi_{j_{L}-1/2,j_{R}}$
\\  \hline
$j_{L}+j_{R}+2$  &$(j_{L},j_{R})$  &(1,0,1)  &0   &$\Phi_{j_{L},j_{R}}$
\\  \hline
$j_{L}+j_{R}+2$  &$(j_{L},j_{R})$  &(0,0,0)  &0   &$\Phi_{j_{L},j_{R}}$
\\  \hline
$j_{L}+j_{R}+2$  &$(j_{L},j_{R})$  &(0,2,0)  &0   &$\Phi_{j_{L},j_{R}}$
\\ \hline
$j_{L}+j_{R}+2$  &$(j_{L}+1/2,j_{R}-1/2)$  &(0,0,2)  &-2  
&$\Phi_{j_{L}+1/2,j_{R}-1/2}$
\\ \hline
$j_{L} +j_{R}+2$  &$(j_{L}+1/2,j_{R}-1/2)$  &(0,1,0)  &-2  
&$\Phi_{j_{L}+1/2,j_{R}-1/2}$
\\ \hline
$j_{L}+j_{R}+2$ &$(j_{L}+1,j_{R}-1)$ & (0,0,0)  & -4 &$\Phi_{j_{L}+1,j_{R}-1}$
\\ \hline
$j_{L}+j_{R}+2$  &$(j_{L}-1/2,j_{R}+1/2)$  &(2,0,0)  &2  
&$\Phi_{j_{L}-1/2,j_{R}+1/2}$
\\  \hline
$j_{L}+j_{R}+2 $ &$(j_{L}-1/2,j_{R}+1/2)$  &(0,1,0)  &2   
&$\Phi_{j_{L}-1/2,j_{R}+1/2}$
\\  \hline
$j_{L}+j_{R}+2$  &$(j_{L}-1,j_{R}+1)$  &(0,0,0)  &4  &$\Phi_{j_{L}-1,j_{R}+1}$
\\  \hline
$j_{L}+j_{R}+5/2$ & $(j_{L}+1/2,j_{R})$  & (1,0,0)  & -1 
&$\Psi_{j_{L}+1/2,j_{R}}$
\\  \hline
$j_{L}+j_{R}+5/2$ & $(j_{L},j_{R}+1/2)$  & (0,0,1)  & 1 
&$\Psi_{j_{L},j_{R}+1/2}$
\\  \hline
$j_{L}+j_{R}+5/2$ & $(j_{L}+1/2,j_{R})$  & (0,1,1)  & -1 
&$\Psi_{j_{L}+1/2,j_{R}}$
\\ \hline
$j_{L}+j_{R}+5/2$  &$(j_{L},j_{R}+1/2)$  &(1,1,0)  &1  
&$\Psi_{j_{L},j_{R}+1/2}$
\\  \hline
$j_{L}+j_{R}+5/2$  &$(j_{L}+1,j_{R}-1/2)$  & (0,0,1)  &-3  
&$\Psi_{j_{L}+1,j_{R}-1/2}$
\\ \hline
$j_{L}+j_{R}+5/2$  &$(j_{L}-1/2,j_{R}+1)$  & (1,0,0)  &3  
&$\Psi_{j_{L}-1/2,j_{R}+1}$
\\ \hline
$j_{L}+j_{R}+3$  &$(j_{L}+1/2,j_{R}+1/2)$ & (0,0,0)  &0  
&$\Phi_{j_{L}+1/2,j_{R}+1/2}$
\\ \hline
$j_{L}+j_{R}+3$  &$(j_{L}+1,j_{R})$ & (0,1,0)  &-2  &$\Phi_{j_{L}+1,j_{R}}$
\\ \hline
$j_{L}+j_{R}+3$  &$(j_{L},j_{R}+1)$ & (0,1,0)  &2  &$\Phi_{j_{L},j_{R}+1}$
\\ \hline
$j_{L}+j_{R}+3 $ &$(j_{L}+1/2,j_{R}+1/2)$ &(1,0,1)  &0  
&$\Phi_{j_{L}+1/2,j_{R}+1/2}$
\\ \hline
$j_{L}+j_{R}+7/2$  &$(j_{L}+1,j_{R}+1/2)$ &(1,0,0)  &-1  
&$\Psi_{j_{L}+1,j_{R}+1/2}$
\\  \hline
$j_{L}+j_{R}+7/2$  &$(j_{L}+1/2,j_{R}+1)$  &(0,0,1)  &1  
&$\Psi_{j_{L}+1/2,j_{R}+1}$
\\  \hline
$j_{L}+j_{R}+4$  &$(j_{L}+1,j_{R}+1)$  &(0,0,0)  &0  &$\Phi_{j_{L}+1,j_{R}+1}$
\\  \hline
\end{tabular}
\end{center}
Table 12. The "massless" supermultiplet
corresponding to the lwv  
$|\Omega \rangle = \xi^{A_1}(1) \xi^{A_2}(1) ... \xi^{A_{2j_{L}}}(1)
\eta^{B_1}(2) \eta^{B_2}(2) ... \eta^{B_{2j_{R}}} (2) |0\rangle =
|\underbrace{\sgenrowbox}_{2j_{L}}, \underbrace{\sgenrowbox}_{2j_{R}} \rangle $.
\vspace{.7cm}

We assume that both $j_{L}, j_{R}$ are either integers or half-integers,
and that $j_{L}, j_{R} \geq 1$.
Otherwise $\Phi$ and $\Psi$ must be interchanged. Note that for 
$j_{L}, j_{R} \geq 1$ this supermultiplet can also be obtained
by tensoring the doubleton supermultiplet from Table 6 with the
doubleton supermultiplet from Table 7. For $j_L=j_R=1$ the above
supermultiplet can also be interpreted as  the $N=4$, spin 4 conformal
 supermultiplet in four dimensions \cite{stelle}.

\section{Implications for CFT/AdS duality}
\setcounter{equation}{0}

We saw in section 8. that there exist doubleton supermultiplets of
ever increasing spin. The  CPT self-conjugate 
irreducible doubleton supermultiplet is that of ${\cal{N}}=4$
super Yang-Mills multiplet in $d=4$. 
One may wonder what the physical meaning of other doubletons
supermultiplets is in light of 
CFT/AdS duality.

In particular, there exist doubleton supermultiplets whose 
spin range is $3/2$. It is also known that there exist
$1/4$ BPS states \cite{ob} in ${\cal{N}}=4$ super Yang-Mills theory in
$d=4$ that correspond to medium long supermultiplets with
spin range $3/2$. 
It would be interesting to find out if these supermultiplets of
$1/4$ BPS states correspond to the medium long doubleton supermultiplets
we found \cite{2mgm}.

We also found doubleton supermultiplets corresponding to ${\cal{N}}=4$
conformal supergravity in $d=4$. Since ${\cal{N}}=4$ super Yang-Mills
theory can be coupled to ${\cal{N}}=4$ conformal supergravity in $d=4$, 
one might wonder if this coupled conformal theory might
describe the dynamics of some higher dimensional theory.
Now, ${\cal{N}}=4$ conformal supergravity has two scalars 
parametrizing the coset space $SU(1,1)/U(1)$.
The $d=10$ IIB supergravity also has two scalars parametrizing 
the coset space $SU(1,1)/U(1)$ \cite{jhs}.
But the original Maldacena's conjecture applies to
the $SL(2,Z)$ invariant sector. This suggests that the full $SL(2,Z)$
covariant dynamics of IIB superstring theory over $S^5$ may be
described in terms of ${\cal{N}}=4$ super Yang-Mills theory
in $d=4$ coupled to ${\cal{N}}=4$ conformal supergravity in
$d=4$, which might point toward a connection between F-theory \cite{vafa} and
CFT/AdS duality.

We should note that the spectrum of scalar fields of the
$S^5$ compactification of $d=10$ IIB supergravity corresponds to 
symmetric tensors of $SO(6)$ ({\bf 20}, {\bf 50},...).
In terms of ${\cal{N}}=4$ superfields,
transforming in the adjoint of $SU(N)$, which represent the
$SU(N)$ gauged 
doubleton supermultiplet \cite{ads2}, the entire spectrum corresponds to
gauge invariant symmetric tensors. We should note that the tensor
product of the doubleton with itself includes the
${\cal{N}}=4$ conformal supergravity multiplet in addition to the 
${\cal{N}}=8$ $AdS_5$ graviton supermultiplet \cite{2mgm}.
 The conformal supergravity multiplet corresponds
to the trace part which is not taken into account by the
original conjecture on CFT/AdS duality.
The fact that the gauge invariant trace component of the
product of two doubletons includes the conformal supergravity
supermultiplet suggests again that the original conjecture of Maldacena
ought to be generalized so as to include the coupling of 
${\cal{N}}=4$ super Yang-Mills in $d=4$ to the
${\cal{N}}=4$ conformal supergravity in $d=4$ as already mentioned above.
The fuller discussion of the implications fo our results to the
CFT/AdS duality, some of which we have already mentioned,
 will be given elsewhere \cite{2mgm}.

CFT/AdS duality has been studied in the literature for fewer 
supersymmetries corresponding to the orbifolding of $S^5$ \cite{ads1}.
Such theories have fewer supersymmetries described by superalgebras
$SU(2,2|k)$ ($k=1,2,3$). Our methods can be extended to these
cases in a straightforward manner.

Finally, one may also wonder how is it possible that the super
 Yang-Mills theory, which comes from the 
open-string sector in $d=10$, captures the dynamics 
of the closed IIB string theory
over $S^5$ which contains gravity. 

Now, the oscillator construction 
of the spectrum of the closed IIB string 
over $S^5$ in terms of the doubleton supermultiplet gives an algebraic
realization of this dynamical relationship between the maximally
supersymmetric Yang-Mills theory and supergravity.
Note also that the oscillator construction
 works very much in parallel with the construction of closed string states in 
terms of open string states in perturbative string theory.

But the oscillator method of constructing the doubleton supermultiplets
(coming from the open string sector) and the graviton supermultiplets
(coming from the closed string sector), which is
obtained by tensoring two doubleton supermultiplets,
should be true non-perturbatively if the CFT/AdS duality
conjecture is indeed correct.
In particular, it should have a deeper dynamical meaning.

It would be important to extend the use of the oscillator method beyond the
calculation of the spectrum to the calculation of the correlation
functions in $d=4$ conformal field theory as was done in 
\cite{witt, ads3}, in order
to uncover the method's true dynamical meaning.
We hope to return to this question in the future. \\

{\bf Acknowledgements}

We would like to thank Shyamoli Chaudhuri, Bernard de Wit, John Schwarz,
Paul Townsend and
especially Juan Maldacena for important discussions.
We thank Sergio Ferrara
for bringing the reference \cite{stelle} to our attention and for pointing
out that some states were missing in the previous version of Table 12.

\section{Appendix}
\setcounter{equation}{0}

The allowed lowest weight vectors (lwv) of $SU(4)$ are given in the
following table

\vspace{.2cm}
\begin{center}
\begin{tabular}{|c|c|c|}
\hline
lwv & SU(4) Dynkin (dim)   &$Y$
\\ \hline
$|0\rangle $ &(0,2,0) (20') & 0
\\ \hline
$\beta^x(1) |0\rangle $  &(0,1,1) ($\bar{20}$)  & -1
\\ \hline
$\alpha^{\mu}(1) |0\rangle$ & (1,1,0) (20)   & 1
\\ \hline
$\beta^x (1) \beta^y (1) |0\rangle $ &(0,1,0) (6) & -2
\\ \hline
$\alpha^{\mu} (1) \alpha^{\nu} (1) |0\rangle $ &(0,1,0) (6)  & 2
\\ \hline
$\beta^x \beta^y \beta^z \beta^w |0\rangle$  &(0,0,0) (1) &-4
\\ \hline
$\alpha^{\mu} \alpha^{\nu} \alpha^{\rho} \alpha^{\lambda}|0\rangle $
& (0,0,0) (1) &4
\\ \hline
$\beta^{(x}(1) \beta^{y)}(2) |0\rangle$  &(0,0,2) ($\bar{10}$) & -2
\\ \hline
$\alpha^{(\mu}(1) \alpha^{\nu)}(2)|0\rangle$ &(2,0,0) (10) & 2
\\ \hline
$\beta^x \beta^y \beta^z |0\rangle$ &(0,0,1) ($\bar{4}$) & -3
\\ \hline
$\alpha^{\mu} \alpha^{\nu} \alpha^{\rho} |0\rangle$ &(1,0,0) (4) &3
\\ \hline
$\alpha^{\mu}(1)\beta^x(2) |0\rangle$ &(1,0,1) (15) &0
\\ \hline
$\alpha^{\mu}(1)\beta^x(2) \beta^y(2)|0\rangle$ &(1,0,0) (4) &-1
\\ \hline
$\alpha^{\mu}(1) \alpha^{\nu}(1) \beta^x(2) |0\rangle$ &(0,0,1) ($\bar{4}$) &1
\\ \hline
$\alpha^{\mu}(1) \alpha^{\nu}(1) \beta^x(2) \beta^y(2) |0\rangle$
&(0,0,0) (1) &0
\\ \hline
\end{tabular}
\end{center}
\vspace{.2cm}

For the decomposition of the supertableaux of $U(m/n)$ in terms of the
tableaux of its even subgroup $U(m)\times U(n)$ we refer to \cite{bbars}. Here
we give a few examples.

\begin{equation}
\begin{array}{rl}
~&~\\
U(m/n) & \supset U(m)\times U(n)\\
~&~\\
\sonebox & = (\onebox,1)+(1,\onebox)\\
~&~\\
\stwobox & = (\twobox,1)+(\onebox,\onebox)+(1,\oneonebox)\\
~&~\\
\soneonebox & = (\oneonebox,1)+(\onebox,\onebox)+(1,\twobox)\\
~&~\\
\stwoonebox & = (\twoonebox,1)+(\twobox,\onebox)+(\onebox,\oneonebox)\\
~&~\\
~& +(1,\twoonebox)+(\oneonebox,\onebox)+(\onebox,\twobox)\\
~&~
\end{array}
\end{equation}

\newpage

\begin{center}
{\bf Erratum \\
 to \\
 " 4D Doubleton Conformal Theories, CPT and IIB
Strings on $AdS_5 \times S^5 $ "}  \\
\end{center}
by {\bf M. G\"{u}naydin, D. Minic and M. Zagermann}, {\it Nucl. Phys.} {\bf B534} (1998) 96-120.
\\

 In the first paragraph of  {\bf section 7} the central charge-like  $U(1)$ is incorrectly identified as the automorphism group $U(1)_Y$ \footnote{We would like to thank Kenneth Intriligator  for informing us of this.}. Therefore, the first part of this paragraph should be replaced by the following: \\

The centrally extended symmetry supergroup of the compactification of
type IIB superstring over the five sphere is the supergroup $SU(2,2|4)$ 
with the even subgroup $SU(2,2) \times SU(4) \times U(1)$, where 
$SU(4)$ is the isometry group of the five sphere [4]. The generator of the Abelian $U(1)$ factor in the even subgroup of $SU(2,2|4)$ commutes with all the 
generators and acts like a central charge. Therefore, $SU(2,2|4)$ is not
a simple Lie superalgebra. By factoring out this Abelian ideal one obtains
a simple Lie superalgebra , denoted as $PSU(2,2|4)$, whose even 
subalgebra is simply $SU(2,2) \times SU(4)$ \footnote[5]{ In [4] the 
symmetry supergroup of the $S^5$ compactification of IIB theory was
denoted as $U(2,2|4)$.}. Both $SU(2,2|4)$ and $PSU(2,2|4)$ have an outer automorphism group $U(1)_Y$ that can be identified with a $U(1)$  subgroup of the $SU(1,1)_{global} \times U(1)_{local}$ symmetry of IIB supergravity in ten dimensions [4]. By orbifolding ....

\end{document}